\documentclass[manuscript]{aastex}

\usepackage{natbib}
\usepackage[utf8]{inputenc}
\usepackage{graphicx}
\usepackage{amsmath}

\begin{document}

\title{Physics Of Eclipsing Binaries. II. \rev{Towards} the Increased Model \rev{Fidelity}.}

\newcommand{\phoebe}{PHOEBE~}
\newcommand{\phoebents}{PHOEBE} 
\newcommand{\Teff}{T_\mathrm{eff}}
\newcommand{\logg}{\log g}
\newcommand{\met}{[\mathrm{M}/\mathrm{H}]}
\newcommand{\vturb}{v_\mathrm{turb}}
\newcommand{\kms}{\mathrm{km}\,\mathrm{s}^{-1}}
\newcommand{\potential}{\mathsf{\Omega}}
\newcommand{\rev}[1]{{#1}}

\author{
  A.~Pr\v sa\altaffilmark{1},
  K.~E.~Conroy\altaffilmark{2,1},
  M.~Horvat\altaffilmark{1,3},
  H.~Pablo\altaffilmark{4},
  A.~Kochoska\altaffilmark{1,3},
  S.~Bloemen\altaffilmark{5},
  J.~Giammarco\altaffilmark{6},
  K.~M.~Hambleton\altaffilmark{1,7} and
  P.~Degroote\altaffilmark{8}
}

\altaffiltext{1}{Villanova University, Dept.~of Astrophysics and Planetary Sciences, 800 E Lancaster Ave, Villanova PA 19085}
\altaffiltext{2}{Vanderbilt University, Dept.~of Physics and Astronomy, 6301 Stevenson Center Ln, Nashville TN, 37235, USA}
\altaffiltext{3}{University of Ljubljana, Dept.~of Physics, Jadranska 19, SI-1000 Ljubljana, Slovenia}
\altaffiltext{4}{Universit\' e de Montr\' eal, Pavillon Roger-Gaudry, 2900, boul.~\' Edouard-Montpetit Montr\' eal QC H3T 1J4}
\altaffiltext{5}{Radboud University Nijmegen, Department of Astrophysics, IMAPP, PO Box 9010, 6500 GL, Nijmegen, The Netherlands}
\altaffiltext{6}{Eastern University, Dept.~of Astronomy and Physics, 1300 Eagle Rd, St.~Davids, PA 19087}
\altaffiltext{7}{Jeremiah Horrocks Institute, University of Central Lancashire, Preston PR1 2HE, UK}
\altaffiltext{8}{KU Leuven, Instituut voor Sterrenkunde, Celestijnenlaan 200D, 3001 Heverlee, Belgium}

\date{Version: 2016-10-15}

\shorttitle{Physics of Eclipsing Binaries}
\shortauthors{Pr\v{s}a et al.}

\date{Received October 1, 2016; accepted mmm dd, yyyy}

\begin{abstract}
The precision of photometric and spectroscopic observations has been systematically improved in the last decade, mostly thanks to space-borne photometric missions and ground-based spectrographs dedicated to finding exoplanets. The field of eclipsing binary stars strongly benefited from this development. Eclipsing binaries serve as critical tools for determining fundamental stellar properties (masses, radii, temperatures and luminosities), yet the models are not capable of reproducing observed data well either because of the missing physics or because of insufficient precision. This led to a predicament where radiative and dynamical effects, insofar buried in noise, started showing up routinely in the data, but were not accounted for in the models. PHOEBE (PHysics Of Eclipsing BinariEs; {\tt http://phoebe-project.org}) is an open source modeling code for computing theoretical light and radial velocity curves that addresses both problems by incorporating missing physics and by increasing the computational fidelity. \rev{In particular, we discuss triangulation as a superior surface discretization algorithm, meshing of rotating single stars, light time travel effect, advanced phase computation, volume conservation in eccentric orbits, and improved computation of local intensity across the stellar surfaces that includes photon-weighted mode, enhanced limb darkening treatment, better reflection treatment and Doppler boosting.} Here we present the concepts on which PHOEBE is built on and proofs of concept that demonstrate the increased model \rev{fidelity}.
\end{abstract}


\section{Introduction}

Eclipsing binary stars (EBs) serve as cornerstones of stellar astrophysics. Their uniquely important role in determining fundamental stellar parameters \citep{popper1980}, distances \citep{pietrzynski2013} and providing rigorous tests for stellar evolution models \citep{torres2010} have been widely appreciated. While the underlying principles that govern the modeling of EBs are simple (Newtonian mechanics and straight-forward geometry considerations), a plethora of complications arises from subtler effects: surface distortions, intensity variations due to gravity darkening, limb darkening, reflection and surface prominences. As the precision of the data increases, so does the number of subtleties that a reliable model needs to take into account.

\begin{figure}[t!]
\includegraphics[width=\textwidth]{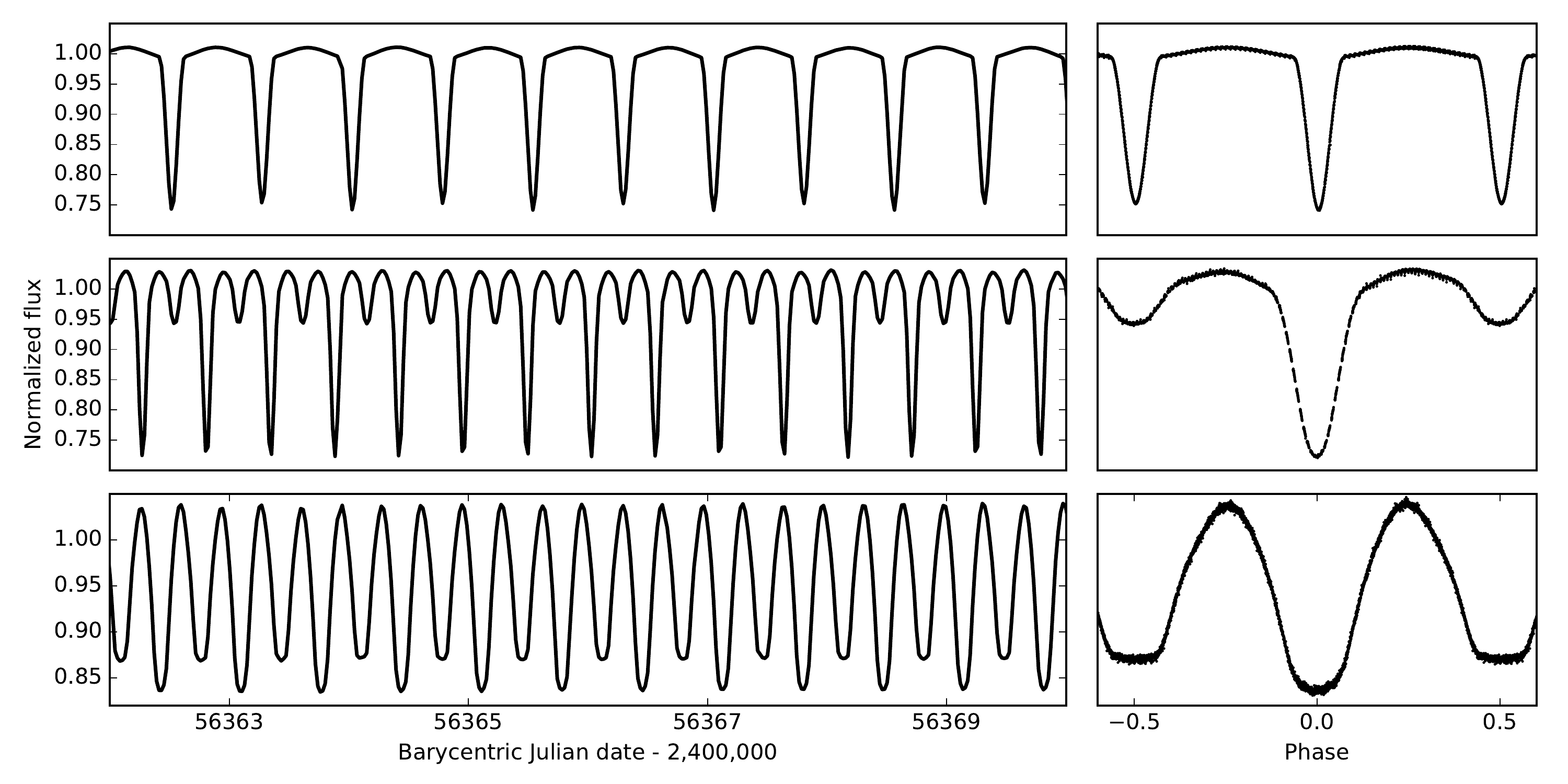} \\
\caption{
\label{fig:keplerlcs}
{\sl Kepler} time series (left) and phased light curves (right) of KIC 5513861 (P=1.51012-d, detached EB), KIC 8074045 (P=0.53638-d, semi-detached EB), and KIC 3127873 (P=0.67146-d, contact EB), top-to-bottom.
}
\end{figure}

This decade has provided us with a range of ground-breaking surveys and missions: {\sl MOST} \citep{walker2003}, {\sl CoRoT} \citep{baglin2003}, {\sl Kepler} \citep{borucki2010}, {\sl Pan-Starrs} \citep{kaiser2002}, {\sl Gaia} \citep{debruijne2012} and {\sl LSST} \citep{tyson2002}, to name just a few. These yielded a vast number of EBs (\citealt{prsa2011a} estimated $\sim$7 million from {\sl LSST} alone), thousands to unprecedented quality (down to $\sim$20 ppm for {\sl Kepler} light curves; cf.~Fig.~\ref{fig:keplerlcs} for several examples of attained precision). We are seeing phenomena that have been up until recently only theorized, but now they are appearing routinely: modern observations provide us with a glimpse into the micromagnitude scale. With such a tremendous boost in both quantity and quality, the tools we use to reduce, analyze and interpret data need to be able to cope with this literal firehose of observations. Yet from all the data currently available, \citet{torres2010} find only $\sim$100 EBs with the uncertainties in the masses and radii smaller than 3\%. For such benchmark objects, this number is several orders of magnitude too low.

Having superb quality data in abundance has clear repercussions on our modeling capability. For the first time we observe astrophysical objects in a near-uninterrupted regime, with uncertainties that dip below 20\,ppm. State-of-the-art models such as the renowned Wilson-Devinney code \citep{wilson1971,wilson1979,wilson2008,wilson2014}, ELC \citep{orosz2000}, and PHOEBE \rev{\citep[paper I]{prsa2005}} are showing systematics in the derived values of fundamental stellar parameters. This is partly because of embedded approximations, partly because of the inconsistent use of physical units and constants (cf.~the IAU 2015 resolution B3; \citealt{prsa2016}) and partly because of the missing and/or inadequate physics built into these models. The residuals \rev{of the best-fit models and observed data} are no longer Gaussian, and we cannot assume that these effects are buried in noise; rather, we need to account for their signatures in light and radial velocity curves explicitly.

Furthermore, the codes provide minimization algorithms that fit model curves to the data. EB data fitting is a highly non-linear problem that suffers from degenerate solutions: the \emph{right} combination of the \emph{wrong} parameters can often fit the observed data as well as the actual solution. The algorithms currently in use, namely Differential Corrections (DC), \citet{powell1964}'s direction set method, \citet{nelder1965}'s Simplex method (NMS), and genetic algorithms \citep{attia2009}, have all met with success, but cannot be run robustly without experienced human intervention, making the tools fully manual.

In this paper we discuss several deficiencies of our models and present advancements in improving the reliability of binary star solutions. \rev{The improvements result from including missing instrumental and astrophysical phenomena in the model explicitly, thus reducing systematical errors in correlated parameters that no longer need to compensate for the missing aspects in the code.} The paper is accompanied by the release of the new version of the open source modeling code \phoebe \rev{2.0, available from {\tt http://phoebe-project.org}, along with extensive documentation and tutorials.}

The layout of the paper is as follows. In Section \ref{s:revision} we provide background on the \phoebe project; Section \ref{s:discretization} introduces the new triangulation scheme for surface discretization; Section \ref{s:dynamics} focuses on dynamical and temporal aspects of the model; Section \ref{s:local_quantities} gives the rationale for computing all local quantities across the mesh; in Section \ref{s:limitations} we discuss current limitations and provide concluding remarks.

\section{Revising the PHOEBE model} \label{s:revision}

The original PHOEBE model \citep{prsa2005} was based on the Wilson-Devinney code \citep[hereafter WD]{wilson1971}. The most important features of the model include: an analytic description of binary star orbits, including apsidal motion and light time travel effect; shape distortion due to tides and (asynchronous) rotation; radiative properties of binary star components, including model atmosphere intensities, gravity darkening, limb darkening and mutual irradiation; and spots, circumbinary attenuation clouds and third light. PHOEBE introduced several scientific and technical extensions: the use of observed spectra as input data, determining individual temperatures by means of color constraining, enhanced reddening treatment, improvements to differential corrections, introduction of the downhill simplex fitting method, the built-in scripting backend, and a graphical user interface.

Since PHOEBE inherited the underlying logic of WD, it also inherited its limitations. These became apparent as ultra-precise photometric data became available: the model could no longer produce adequate fits to light curves at {\sl Kepler}'s level of precision. In addition, a wealth of triple and multiple stellar systems have been discovered that exhibit eclipse timing variations due to the light time travel effect and dynamical effects, circumbinary planets and multiply eclipsing systems, all of which render the old model unsuitable. To overcome these inherited limitations, the new version of PHOEBE presented here has been rewritten from scratch. We present all novel aspects of the model, benchmarks that prompted a redesign, and comparison between original and revised models. The rewrite of the code was initiated in 2011 but was limited by the lack of funding; the current release has been made possible by the NSF support \#1517474, which we gratefully acknowledge. 

The version of \phoebe presented here is \rev{2.0}.  The code is \emph{not} backwards-compatible to \phoebe 1.0 because of the complete redesign of the backend, but original \phoebe parameter files can be imported into the new version seamlessly.  Versioning will follow the common semantic versioning syntax (major.minor.patch) with any critical bug fixes noted with an increment in the patch number (i.e.~2.0.1, 2.0.2), any new features noted with an increment in the minor number (i.e.~2.1.0, 2.2.0) and usually accompanied by a publication describing the new features, and any major new releases noted by an increment in the major number (i.e.~2.0.0, 3.0.0). \rev{If the patch number is 0, it can also be omitted (hence \phoebe 2.0 instead of \phoebe 2.0.0).} Going forward with the \rev{2.x versions} of PHOEBE, all efforts will be made to maintain backwards compatibility, with any unavoidable changes listed clearly in the release notes and requiring a new release.

\phoebe is released as open source, under the General Public License. It is written in C (backend) and python (frontend). \phoebe is available for download from {\tt http://phoebe-project.org}.

\section{Discretization of implicit surfaces} \label{s:discretization}

\begin{figure}[t!]
\includegraphics[width=\textwidth]{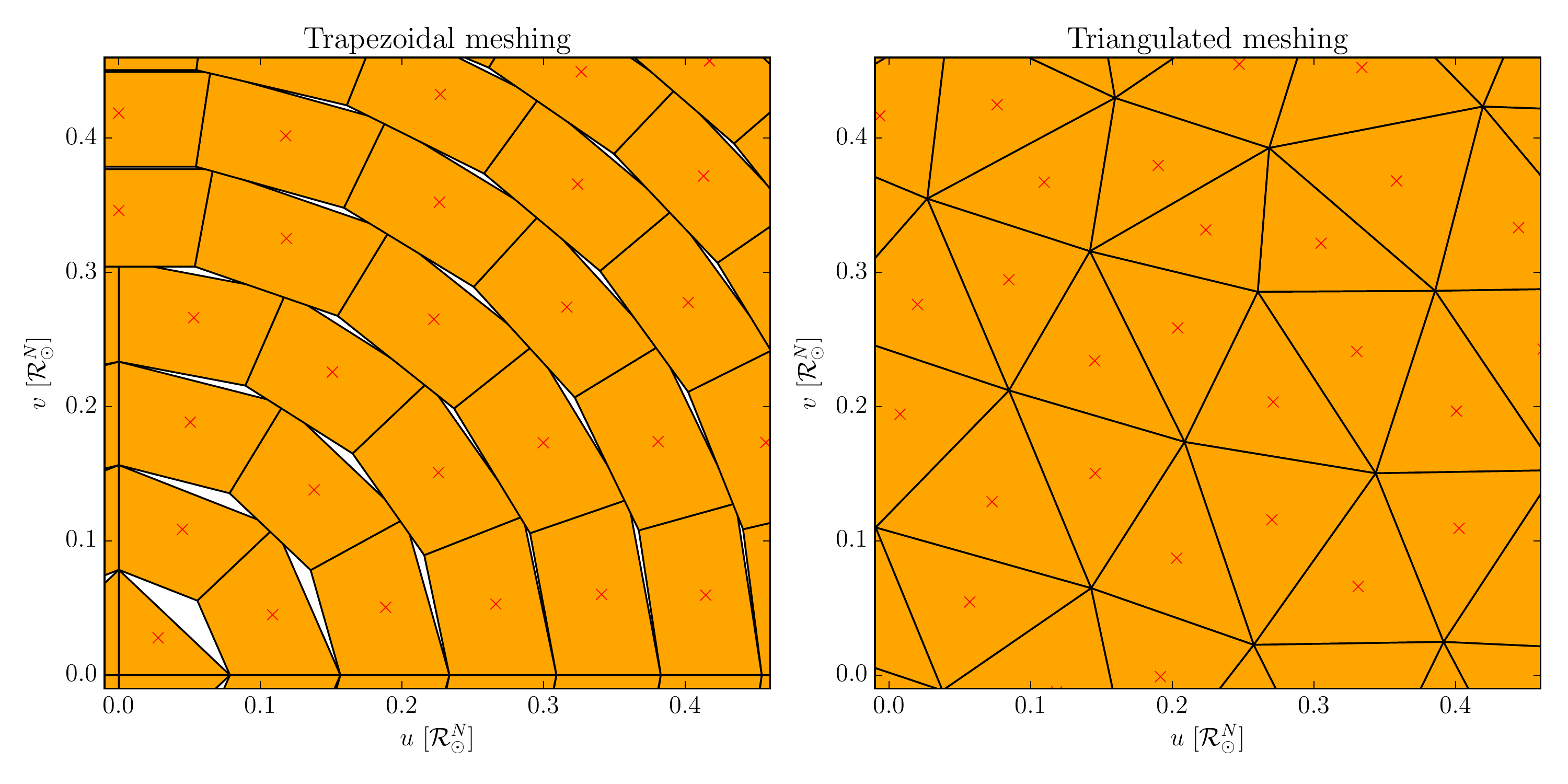}
\caption{
\label{fig:mesh_comparison}
Mesh comparison between the trapezoidal discretization using trapezoidal elements (left) and triangulation (right). The crosses denote surface element centers. Both parts correspond to the same region on the star. The issues with trapezoidal discretization are disconnected, overlapping meshes with holes, and obvious ``seams'' across the surface. All these cause systematic effects in computed fluxes.}
\end{figure}

Computing the amount of flux received from an EB becomes a non-trivial task as soon as there is even a marginal surface distortion of the EB components. The employed generalized Roche geometry provides us with that distortion, and a range of related phenomena is then taken into account: surface brightness variation due to gravity darkening, limb darkening and reflection, ellipsoidal variations due to the changing cross-section size that faces the observer, spots coming in and out of view, etc. The dynamical aspects then render the two components onto their respective orbits and a numerical integration is done over the visible surfaces, determined by eclipses and visible cross-sections, to compute the total flux. It is thus imperative that local conditions across the surfaces of both components be approximated as well as possible. To achieve this, the surfaces were traditionally discretized uniformly along co-latitude and longitude into planar, trapezoidal elements for which we assume uniform local conditions. The most significant problem with this strategy is that the surface is approximated by a \emph{disconnected}, \emph{self-overlapping} mesh that does not cover the surface completely. The left panel in Fig.~\ref{fig:mesh_comparison} displays a part of the star near a pole, a region where these problems are particularly exacerbated. To combat this deficiency, surface element areas are not computed from the trapezoids themselves but from the theoretical surface element differentials, $d\sigma = \frac{1}{\cos \beta} \rho^2 \sin \theta d\theta d\phi$, where $\rho$, $\theta$ and $\phi$ are spherical coordinates, and $\beta$ is the angle between the surface normal and the radius vector. The problems become more pronounced as the components become distorted, and especially when we deal with ultra-precise photometric data, since a uniform sampling in co-latitude and longitude implies that the sampling in terms of surface coverage is not uniform.

To overcome this problem, we replaced trapezoidal surface elements with triangles. The triangulation of implicit surfaces is a long standing problem in computational physics. \citet{hartmann1998} presents a \emph{marching method}: an algorithm that can discretize any implicit surface into a set of near-equilateral triangles, so that every element on the surface has approximately the same area, irrespective of the position or the amount of surface distortion. The right panel in Fig.~\ref{fig:mesh_comparison} demonstrates this: the polar region of the star has no inherent symmetry that would impose ``seams'' across the surface, or surface elements of notably different sizes. In effect, any systematics that may arise from surface discretization will be minimized. A mesh generated this way is computationally tractable and it alleviates the problems with holes or overlaps between trapezoidal elements. An added benefit is that we can discretize any implicit geometric body (i.e.~accretion disks, disintegrating planets, alien megastructures) and model any transient phenomena within the same formalism.

Triangulation by itself does not solve the problem of unequal surface areas of the mesh and of the isosurface that the mesh approximates. This leads to a similar problem that trapezoidal discretization suffers from: while a trapezoidal mesh overestimates the area because surface element centers are on the isosurface, triangulation underestimates the surface area because the vertices are on the isosurface. To correct for this, we offset all surface elements such that (1) the total surface area of the mesh exactly equals the analytical surface area of the isosurface, and (2) each surface element approximates the local isosurface optimally, i.e. it is positioned in a way that exactly balances isosurface deviations due to curvature below and above the surface element. The details of the mesh offsetting scheme are given in Appendix \ref{appendix:offsetting}.

PHOEBE currently supports two types of surface potentials: the Roche model and the rotating single star model. In general, though, any surface that can be described by an implicit equation can be readily incorporated into the code.

The generalized Roche lobe \citep{wilson1979} is defined by:
\begin{equation}
  \potential_{\rm Roche}({\bf r}; q, F, \delta) =
  \frac{1}{\|{\bf r}\|}
  + q \left (
    \frac{1}{\sqrt{(x-\delta)^2 + y^2 + z^2}} - \frac{x}{\delta^2}
  \right)  + 
  \frac{1}{2} (1 + q) F^2(x^2 + y^2),
  \label{eq:roche_potential}
\end{equation}
expressed in Cartesian coordinates $x$, $y$ and $z$, with the radius vector denoted by ${\bf r}=(x,y,z)$; $q \equiv M_2/M_1$ is mass ratio, $F \equiv P_\mathrm{rot}/P_\mathrm{orb}$ is synchronicity parameter and $\delta \equiv d/a$ is instantaneous separation between the stars relative to the semi-major axis $a$.
 
\begin{figure}[t!]
\centering
\includegraphics[width=0.475\textwidth]{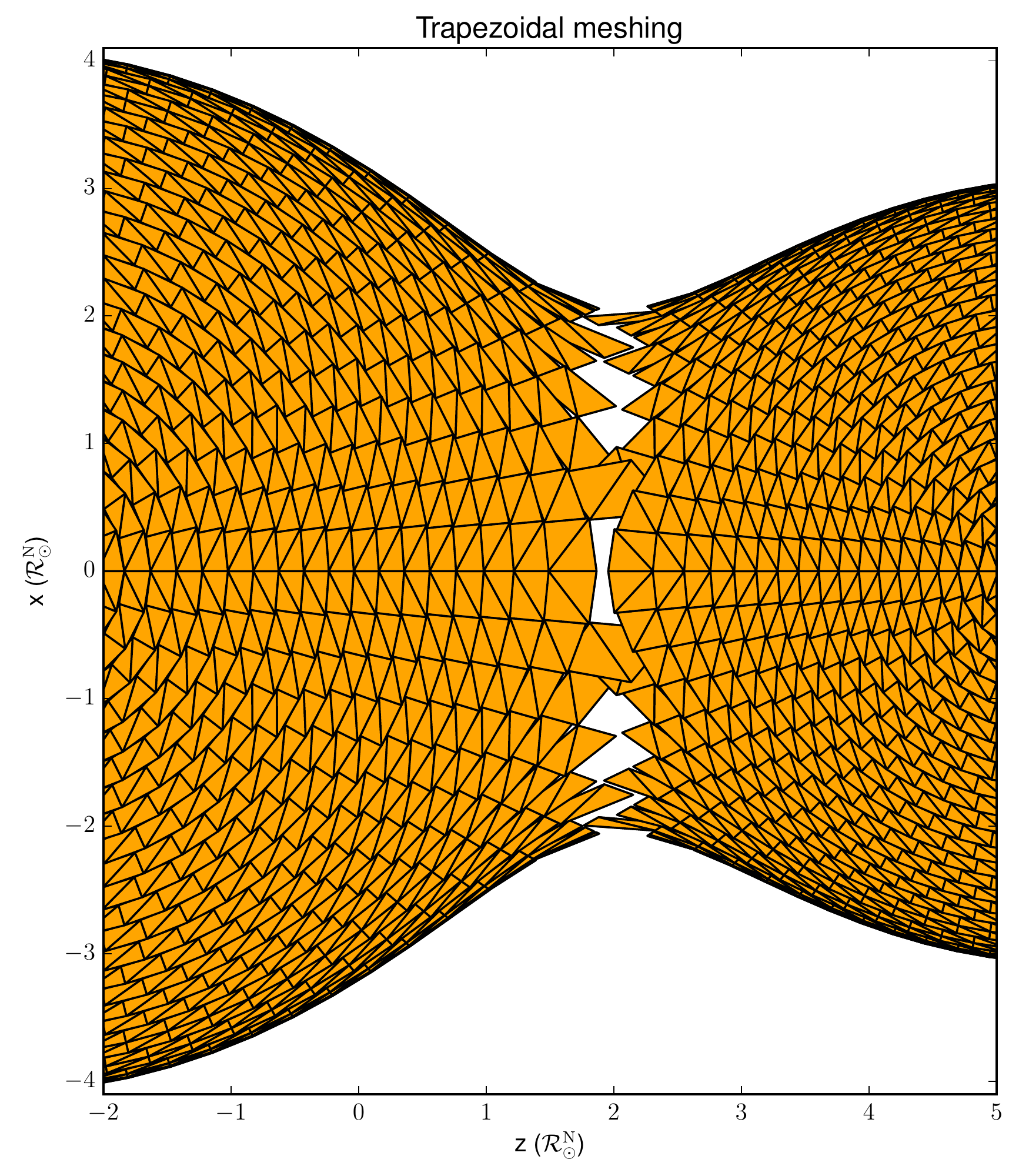}
\includegraphics[width=0.475\textwidth]{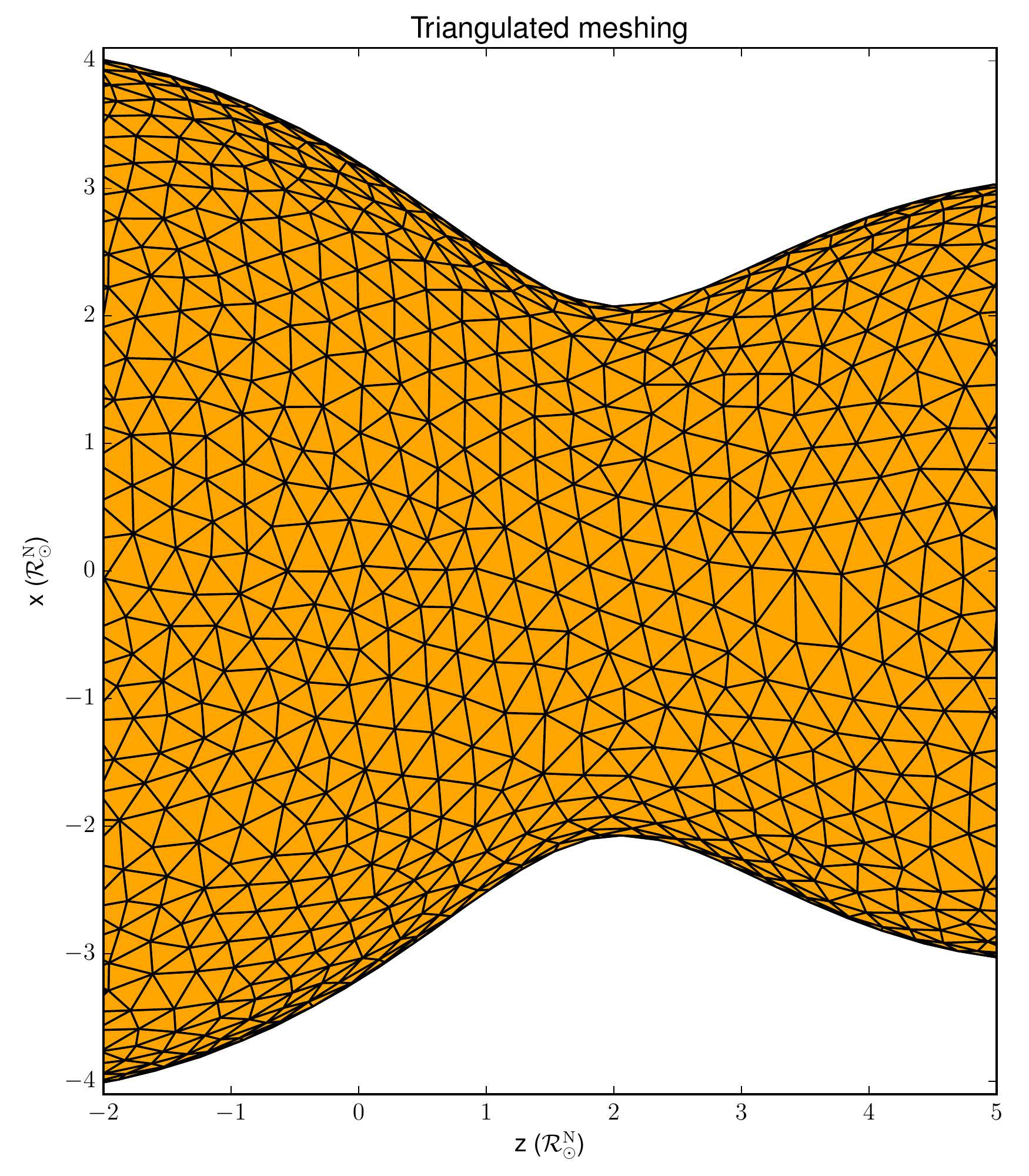}
\caption{\rev{Comparison of the neck region coverage in a trapezoidal (left) and triangulated mesh (right) of a contact system with mass ratio $q = 0.5$ and fillout factor $FF = 0.9$. The trapezoidal mesh is represented by its current implementation in \phoebe, in which each trapezoid is split into two triangles for plotting, but retains its center, where all local quantities are computed.}}
\label{fig:oc_meshes}
\end{figure}

A significant difference between the marching and trapezoidal meshes arises in contact\footnote{The original PHOEBE version, following the convention of the Wilson-Devinney nomenclature, used the term ``overcontact''; here and throughout PHOEBE 2.0 we use the term ``contact'' as it avoids confusion with the outer lobe overflow and mass loss through $L_2$ and/or $L_3$ points, which is what ``overcontact'' might imply. See \citet{rucinski1997} and \citet{wilson2001} for an in-depth discussion of this issue.} binary systems computed with the Roche model. The trapezoidal method tends to diverge near the neck of contact envelopes, which requires the meshing of the surface to account for each star separately and ''glue`` them together at the neck \citep{wilson1979}. This can often lead to poor coverage of the neck region, particularly for systems with a small mass ratio \rev{and large fillout factor}. The marching method, on the other hand, computes the entire contact binary mesh as one surface, with complete coverage of the neck region. \rev{Examples of a trapezoidal and triangulated mesh for a contact system are given in Fig.~\ref{fig:oc_meshes}.}

The potential of an isolated star rotating around the $z$-axis is given by:
\begin{equation}
  \potential_{\rm rot}({\bf r}; \omega) = \frac{1}{\|{\bf r}\|} + \frac{1}{2} \omega^2 (x^2 + y^2),
  \label{eq:rotstar_potential}  
\end{equation}
where $\omega$ is the angular velocity of rotation.
Support for other surface potentials will follow shortly, most notably for rotating and non-rotating spheres (designed for exoplanets), misaligned Roche potentials, general gravitational potentials (useful for multiple stellar systems) and accretion disk potentials (Horvat et al., in preparation).

\section{Dynamical aspects} \label{s:dynamics}

Motions of two celestial bodies around the mutual center of mass are governed by Kepler's equations \citep{goldstein1980}. The orbits depend on the semi-major axis $a$, orbital period $P_0$, mass ratio $q$, eccentricity $e$ and systemic velocity $v_\gamma$; their orientation is given by inclination $i$, argument of periastron $\omega$ and longitude of the ascending node $\Omega$. While this formalism describes dynamical positions of the bodies in their orbits, it does not describe the position where the observer \emph{sees} the bodies due to the finite speed of light. This is all the more important when there are additional objects in the system because the systemic velocity of the binary is no longer constant. 

To address this, \phoebe defines the time reference $t_0$ with respect to the barycenter of the entire system, with a provision that the barycenter can move at a constant speed through space. The Light Travel Time Effect (LTTE) causes bodies closer to the observer w.r.t.~barycenter to appear at an advanced point in orbit, while bodies farther from the observer w.r.t.~barycenter appear retarded. In addition to the LTTE effect, perturbations of orbital elements (period change, apsidal motion, mass transfer, Kozai cycles, \dots) also affect timing \citep{borkovits2007}. \phoebe optionally accounts for these effects by computing all positions w.r.t.~barycenter and then iteratively advancing or retarding the objects on their orbits by the amount consistent with their radial distance from the barycenter.

Support for hierarchically-nested Keplerian orbits as well as dynamics being driven by an N-body integrator, particularly applicable to triple and higher-order systems, is being developed (Conroy et al., in preparation).

\subsection{Phase computation}

Periodicity in eclipsing binary observables due to orbital dynamics is a backbone of modeling. All computations in PHOEBE are done in time-space, but it is often convenient to represent observables in phase-space as well. It is thus important to set a convention for phase. It is common practice to set the reference point $t_0$ at the time of superior conjunction (primary eclipse), but this is not a suitable convention because of apsidal motion (rotation of the line of apsides). Apsidal motion causes excursions of \emph{both} conjunctions in the opposite directions with uniformly advancing time. Keeping a reference point at conjunction would cause us to measure an anomalous orbital period. In consequence, whatever reference point is chosen, it should not be linked to any particular point on the orbit (i.e.~conjunction, periastron passage, etc). Instead, we define phase of conjunction w.r.t.~the mean anomaly:
\begin{equation}
    \Phi_C = \frac{M_C+\omega}{2\pi} - \frac 14,
\end{equation}
where mean anomaly $M_C$ is connected to the true anomaly via the familiar Kepler equations, and zero point in phase is set by convention to coincide with phase of conjunction when argument of periastron $\omega = \pi/2$. This is chosen because the true anomaly (being measured from periastron) is $\upsilon_C = \pi/2 - \omega$ at superior conjunction. PHOEBE provides phases on the [-0.5, 0.5] interval.

To transform time to phase while allowing the period to vary linearly (i.e. $\dot P \neq 0$), we solve the following differential equation:
\begin{equation}
  d\Phi = \mathrm{mod} \left[ \frac{dt}{P(t)} \right] \Rightarrow \Phi(t) = \mathrm{mod} \left[ \int \frac{dt}{P_0 + \dot P (t - t_0)} \right] = \mathrm{mod} \left[ \frac{1}{\dot P} \log [P_0 + \dot P (t-t_0)] \right],
\end{equation}
where $P_0$ is period at reference time $t_0$, and $\mathrm{mod}$ is the modulo operation that maps its argument to the $[-0.5, 0.5]$ interval. For $\dot P = 0$, $\Phi(t)$ assumes a familiar expression:
\begin{equation}
  \Phi(t) = \mathrm{mod} \left[ \frac{t-t_0}{P_0} \right].
\end{equation}

Note also that phase shift is a parameter that applies only to synthetic curves; it is used to reconcile the horizontal offset between data and the model when $t_0$ is kept at a geometric reference point, i.e.~superior conjunction or argument of periastron. Phase shift does not affect phasing of observations.

\subsection{Volume conservation in eccentric orbits}\label{sec:volume_conservation}

A non-trivial assumption that partly follows from physical insight and partly from observations is stellar volume conservation in eccentric orbits. One of the fundamental assumptions of the Roche model is that a star adapts to the instantaneous force field on a timescale that is significantly shorter than the orbital period. In other words, a star will instantly change its shape to adapt to the equipotential that bounds it. In the case of eccentric orbits, however, equipotential lobes can change their volumes substantially. If a star adapted to this instantaneous equipotential at every point in its orbit, that would mean that it needs to rarefy or compress in order to fill the volume of the equipotential. In that case, some of the heat generated during compression would be irradiated away and thus not recovered during subsequent expansion. This energy would need to be replaced from the energy stored in orbit, leading to rapid circularization since even a tiny energy loss compounds quickly over many orbits. In effect, there would be essentially no eccentric short period binaries in the sky. This would also significantly impact the efficiency of Kozai cycles, which tighten the orbits on timescales orders of magnitude longer than the orbital period.

Observationally, if the volumes of stars changed on eccentric orbits, we would expect to see this effect in the measured radii of eclipsing binaries. However, eclipse durations are consistent with near-constant radii and, thus, near-constant volumes. Even the red giant and supergiant stars that would arguably be most susceptible to volume change don't exhibit any signs of an appreciable change in radius as a function of orbital phase. The star does not adapt to the equipotential but, instead, conserves its volume and fills a \emph{different} equipotential to that at periastron. We thus conclude that volume conservation is a safe assumption with strong footing in both theoretical insight and observations.

That said, there is a particularly well suited test-bed for studying the validity of volume conservation in the limit of high eccentricities: highly eccentric ellipsoidal variables (also known as \emph{heartbeat} stars; \citealt{thompson2012}) that host red giants. Their components at periastron are separated by only a few stellar radii, giving rise to significant tidal forces that distort stellar surfaces. A total of 28 giants in heartbeat stars have been reported to date \citep{nicholls2012,gaulme2013,gaulme2014,beck2014,richardson2016}; of these, there are 9 for which the authors developed a full model \citep{nicholls2012,beck2014,richardson2016}, all of them under the constant volume assumption. Yet these are the prime candidates where small deviations from constant volume are plausible. Thus, even though the default mode of operation is to keep the volume constant, we allow the volume to change by a prescribed amount so that we can rigorously study the robustness of the volume conservation assumption. This will be the topic of Hambleton et al., in preparation.

To handle volume conservation within \phoebents, the meshes are first built according to the equipotentials defined at periastron and their analytic volumes are computed (see Appendix \ref{appendix:volume_surfacearea}).  When placed in orbit at any given time, the current instantaneous equipotential is then determined to be the equipotential that will result in this same volume.

\section{Sampling local quantities}\label{s:local_quantities}

In PHOEBE 1, the trapezoidal surface elements were placed so that the center of each element is on the equipotential surface.  When using the marching method, the vertices of the triangles are placed on the equipotential.  This allows us to sample all local physical quantities at the vertices rather than the centers of the surface elements.  The physical quantity assigned to each fully visible triangle (partially visible triangles are discussed in Section \ref{eclipse detection}) is the arithmetic mean of the values at each of the vertices. This is a linear approximation, i.e.~we assume that the variation of each physical quantity across each discretized surface element is linear.  It is therefore necessary to sample the surface with a sufficiently fine grid so that this approximation is accurate.

Building the mesh so that either the vertices or the centers of the elements are placed on the equipotential both suffer from the projected surface being either smaller or larger than the true surface, respectively (cf.~discussion in Section \ref{s:discretization}).  Although the local quantities are defined on the surface appropriately, the shadow created by the mesh is either smaller or larger than the equipotential surface, resulting in systematically offset times of ingress and egress. To correct this, we calculate two meshes: one in which the vertices are placed on the surface, and the other that is adjusted to the correct shadow size. The first mesh is used for all physical quantities, while the second mesh is used for all eclipse determination and geometric quantities, such as surface area of each element. As discussed in Section \ref{s:discretization}, the second mesh is computed by offsetting each vertex along its normal such that the surface area of the mesh matches the expected theoretical surface area of the Roche equipotential.  This operation depends on a local curvature of the mesh (see Appendix \ref{appendix:offsetting} for details).

\subsection{Effective Temperatures}

Local effective temperature distribution across the stellar disk depends on many factors, but predominantly on the tidal and rotational distortion of the stellar surface. We model this distribution as a simple power law where local effective temperatures are computed for each vertex from the \emph{polar} effective temperature: $\Teff = T_\mathrm{pole} \mathcal{G}^{1/4}$, where the gravity darkening coefficient $\mathcal{G} = \left( g / g_\mathrm{pole} \right)^{\beta}$ is defined for each vertex, $\beta$ is the gravity darkening parameter of the model and $g$ is local surface gravity acceleration per vertex. The polar temperature is computed from the mean effective temperature via:
$$
  T_\mathrm{pole} = T_\mathrm{eff, mean} \left( \frac{\sum_i A_i }{\sum_i \mathcal{G}_i A_i } \right)^{1/4},
$$
where $A$ is the area of each surface element and $T_\mathrm{eff, mean}$ is the model parameter and a global property of the star.

The initial support for contact binaries in \phoebe 2.0 retains the current approach to their treatment, with modifications to adapt it to marching meshes. In the trapezoidal approach, a boundary plane that separates the two components at the position of the minimum neck radius is computed \citep{wilson1976}. The centers of the trapezoids are then split with respect to the side of the plane they belong to and the fractional area of each trapezoid that intersects the boundary plane is retained as the area of its corresponding center. The fractional area of a trapezoid that partially belongs to one component, while its center belongs to the other, is added to the area of the center nearest in longitude. \rev{The fractional area assignment ensures that the mesh is well connected and deals with the overlaps between the trapezoids of the two components, but does not solve the issue of diverging centers, which may leave gaps in the mesh.}

Since all surface quantities of a marching mesh are computed in the triangle vertices, the split between components in a marching mesh of a contact binary is applied to vertices as opposed to centers. \rev{Whether a triangle and all its vertices belong to the primary or secondary component is determined based on the position of the triangle center, equivalent to the trapezoidal approach. All of the triangle vertices are then marked as belonging to one component, despite of the fact that some might cross the boundary plane. The components of the contact envelope have surface temperature distributions computed via their respective polar temperatures. For components with different temperatures this introduces unphysical discontinuities in the neck area, as neighboring triangles that belong to distinct components can have very different temperatures. This approach is thus mainly suitable for systems close to thermal equilibrium and we advise against its use for systems not in thermal equilibrium until a smooth temperature variation has been implemented (see Section \ref{s:limitations}; Kochoska et al., in preparation).}

PHOEBE supports a simplistic spot model inherited from PHOEBE 1 (by way of WD).  Spots are defined by their colatitude, longitude, angular radius, and a temperature factor.  Any vertex that falls within the boundary of this circular spot region acquires a temperature that equals the product of the intrinsic effective temperature and the temperature factor. Support for more complex spot treatment, including surface migration and size change, is being planned for the near future.

\subsection{Model atmospheres}

\begin{figure}[t!]
\begin{center}
\includegraphics[width=\textwidth]{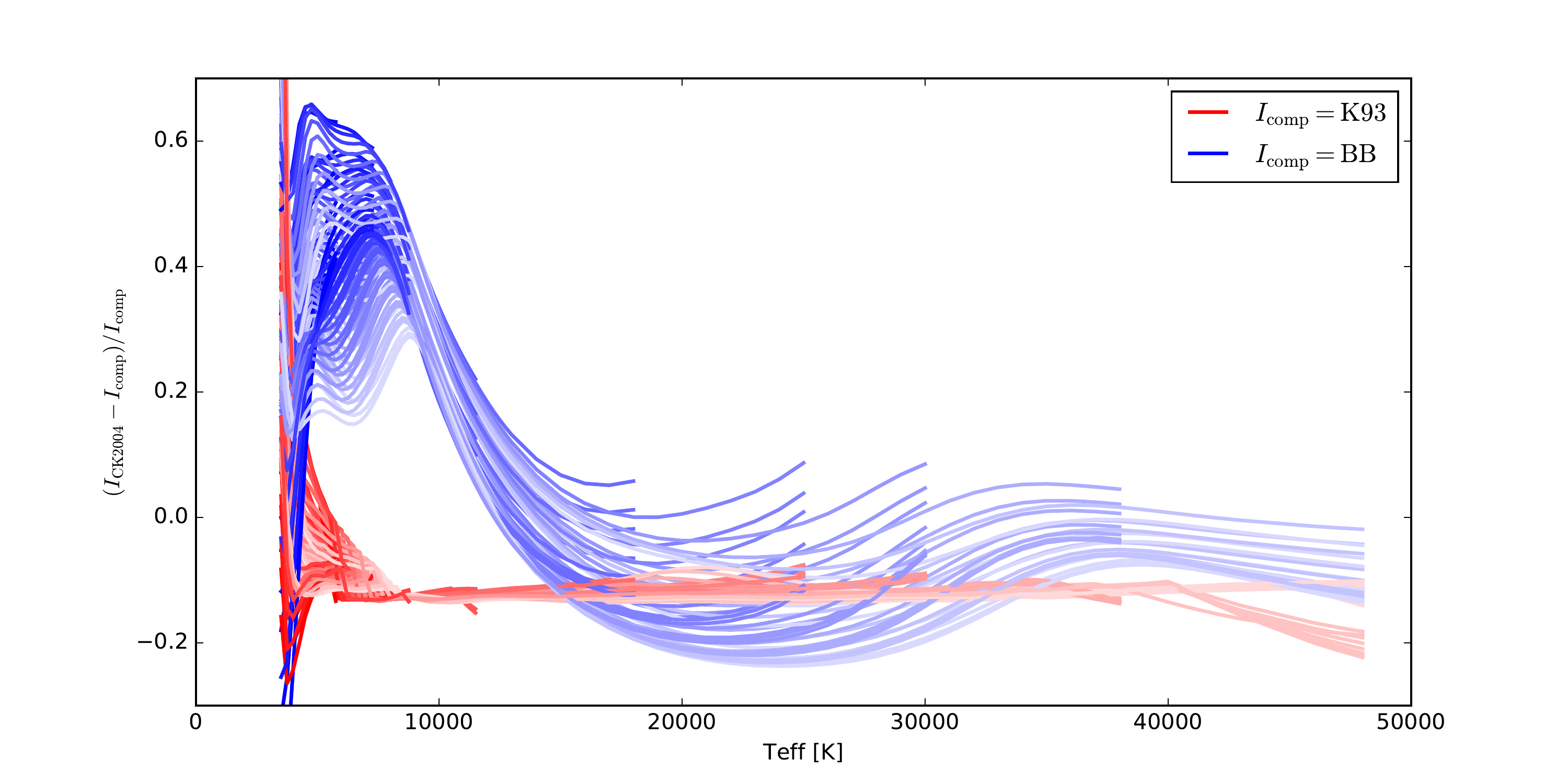} \\
\end{center}
\caption{
\label{fig:atmcomp}
\rev{Relative difference between the normal emergent passband intensities computed by the \citet{castelli2004} model atmospheres, the original \citet{kurucz1993} model atmospheres (red), and blackbody atmospheres (blue). Color tones correspond to the value of $\log g$, where the darkest tone corresponds to the lowest $\log g$ and the lightest tone corresponds to highest $\log g$ value. Each line corresponds to a combination of $\log g$ and metal abundance. Model atmosphere flux differences are most notable on the low temperature end (below $\sim$8000\,K), in excess of 200\% for M stars. Blackbody fluxes deviate across the entire temperature range.}
}
\end{figure}

Local normal emergent intensity is a complex function of local thermodynamical and hydrodynamical properties, most notably the local effective temperature, surface gravity, abundance of heavy elements, rotation and microturbulent velocity, but it also depends on extrinsic effects, such as Doppler boosting, interstellar extinction, extraneous light contamination, etc. The overly simplified case is to consider stars as blackbody radiators. In that case the local normal emergent intensity depends solely on the local effective temperature, via Planck's law. A more realistic treatment necessitates the use of model atmospheres (i.e.~\citealt{hubeny1995}, \citealt{hauschildt2003}, \citealt{castelli2004}). Model atmospheres provide intensity as a function of the above mentioned quantities, and we need to assign an appropriate model atmosphere to each component in the modeled system, where the choice depends considerably on the type of star and on the intended use. For example, plane-parallel model atmospheres might be perfectly adequate for Sun-like stars, whereas one would likely opt for spherical, non-local thermodynamic equilibrium (NLTE) models for hot O- and B-type stars. Similarly, a detailed, full-featured model atmosphere is required for comparing syntethic spectral energy distribution (SED) with observed spectral lines, while area-preserving parameters such as microturbulent velocity may be marginalized out if we are integrating SEDs to obtain local emergent passband intensities. \rev{Fig.~\ref{fig:atmcomp} depicts a comparison between the \citet{castelli2004}, original \citet{kurucz1993} and blackbody atmospheres for a range of $\log g$ and abundance values (denoted by different tones). The difference between the two model atmospheres at higher effective temperatures is $\sim$12\%, and in excess of $\sim$200\% at the lowest temperatures (not depicted in the figure to highlight the variation details rather than the overall amplitude). The flatness beyond $\sim$8000\,K implies that \emph{relative} flux computations will vary to the extent of the deviation from the flat curve, but the \emph{absolute} flux computations will differ by as much as $\sim$15\%.}

Model atmospheres in PHOEBE are used to (1) compute normal emergent passband intensity, i.e. local intensity perpendicular to the surface of the star integrated over a given passband response function; (2) compute center-to-limb variation in intensity, also known as limb darkening; (3) compute the increase in intensity due to reflection effect; and (4) compute Doppler boosting due to kinematic properties of stellar surfaces. In the near future PHOEBE will also be able to synthesize theoretical SEDs and compare them with observed spectra. We provide some basic considerations for each use case next.

\subsubsection{Passband intensity}

Local normal emergent \emph{passband} intensity $I_\mathrm{pb}$ is obtained by multiplying the SED $\mathcal S (\lambda) \equiv dI(\lambda)/d\lambda$ in the direction perpendicular to the surface ($\mu \equiv \cos \theta = 1$) with the passband transmission function $\mathcal P (\lambda)$ and integrating over wavelength:
\begin{equation} \label{eq:passband_intensity}
I_\mathrm{pb} (\Teff, \logg, \met, \vturb, \dots) = \int_\lambda \mathcal S (\lambda; \Teff, \logg, \met, \vturb, \dots) \mathcal P(\lambda) d\lambda.
\end{equation}
This is a time-consuming operation considering that for each local circumstance we need to synthesize the corresponding model atmosphere, multiply it with $\mathcal P(\lambda)$ and integrate it. To alleviate this computational burden, for each passband we store the integrated values of $I_\mathrm{pb}$ for an extensive grid of $\Teff$, $\logg$ and $\met$ values based on the \citet{castelli2004} models. We then interpolate in these three quantities to get an accurate estimate of $I_\mathrm{pb}$. 

\begin{figure}[t!]
\begin{center}
\includegraphics[width=\textwidth]{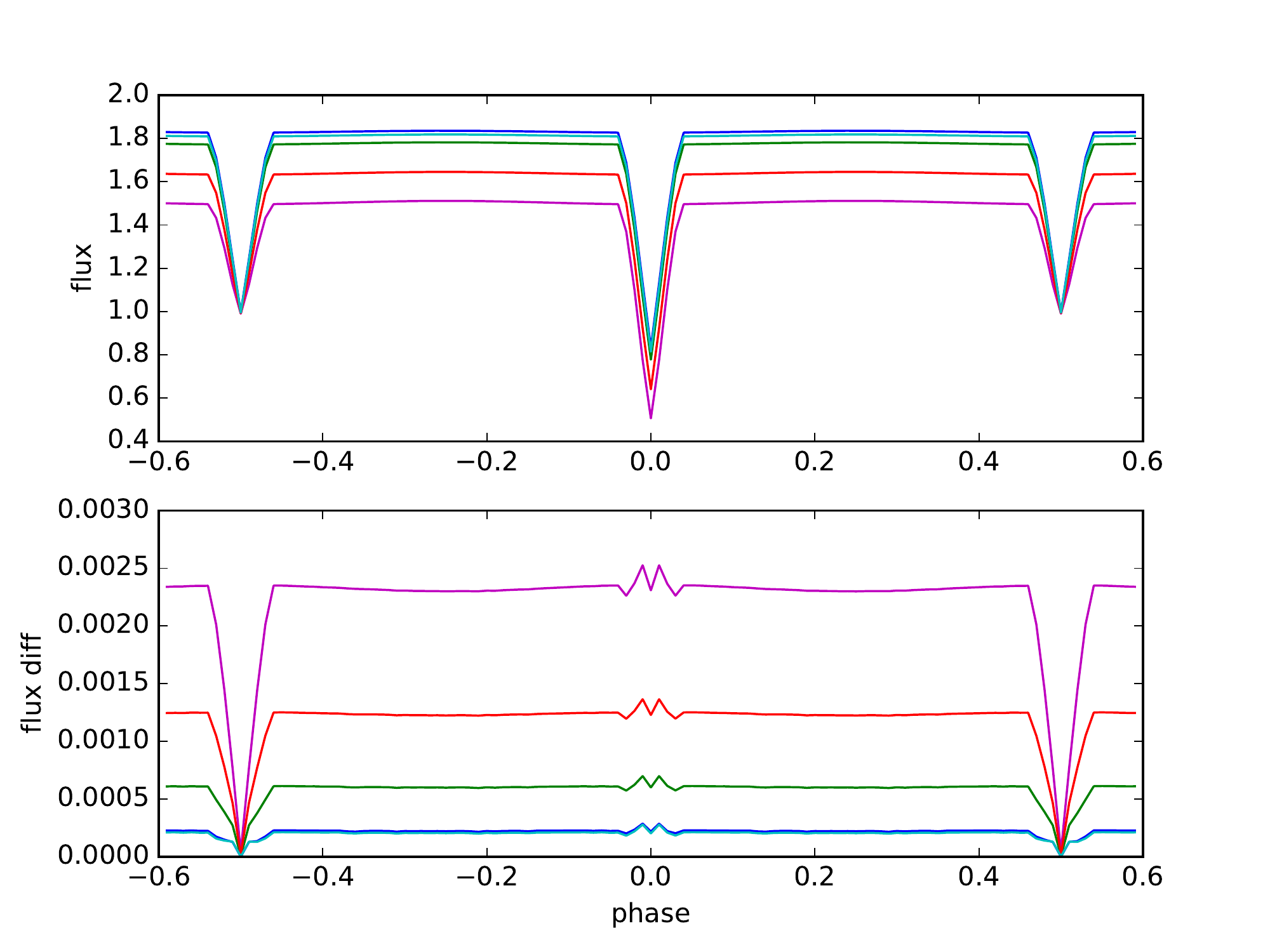}
\caption{
\label{fig:intens_weighting}
Light curves for detached binaries with a temperature ratio of 0.9 \rev{in the Johnson V passband} as a function of primary temperature ($T_{\mathrm{eff},1} = 5000$\,K (purple), 7500\,K (red), 10000\,K (green), 12500\,K (cyan) and 15000\,K (blue).  The lower panel shows the difference caused by weighting the intensities by photon count vs energy.
}
\end{center}
\end{figure}

Depending on the detector type used, Eq.~(\ref{eq:passband_intensity}) may not be appropriate. It implicitly assumes that our detector is bolometric in the sense that it measures intensity by accumulating energy, i.e.~that intensity is \emph{energy-weighted}. This is appropriate for observations that are flux-calibrated using standard stars. In contrast, photon-counting devices need to be \emph{photon count-weighted}. Considering that the photon energy is $E_\nu = hc/\lambda$, Eq.~(\ref{eq:passband_intensity}) can be rewritten as:
\begin{equation} \label{eq:passband_intensity_photon}
I_\mathrm{pb}' (\Teff, \logg, \met, \vturb, \dots) = \frac{1}{hc} \int_\lambda \lambda \, \mathcal S (\lambda; \Teff, \logg, \met, \vturb, \dots) \mathcal P(\lambda) d\lambda,
\end{equation}
where $I_\mathrm{pb}'$ can still be considered intensity, but in units of photon counts. This version is more appropriate for observations where counts are not calibrated (including instrumental and differential photometry, including {\sl Kepler} data) and passbands are wide \citep{bessell2005, bessell2012, casagrande2014, brown2016}. \phoebe supports both regimes and tabulates $I_\mathrm{pb}$ values for energy-weighted and count-weighted intensities. The difference for solar-type stars amounts up to 1\%.  Figure \ref{fig:intens_weighting} shows the difference between these as a function of the temperature of the stars. The effect is larger for \rev{cool} stars \rev{(purple line)} \rev{and is also dependent on the width of the passband}.

\begin{figure}[t!]
\begin{center}
\includegraphics[width=\textwidth]{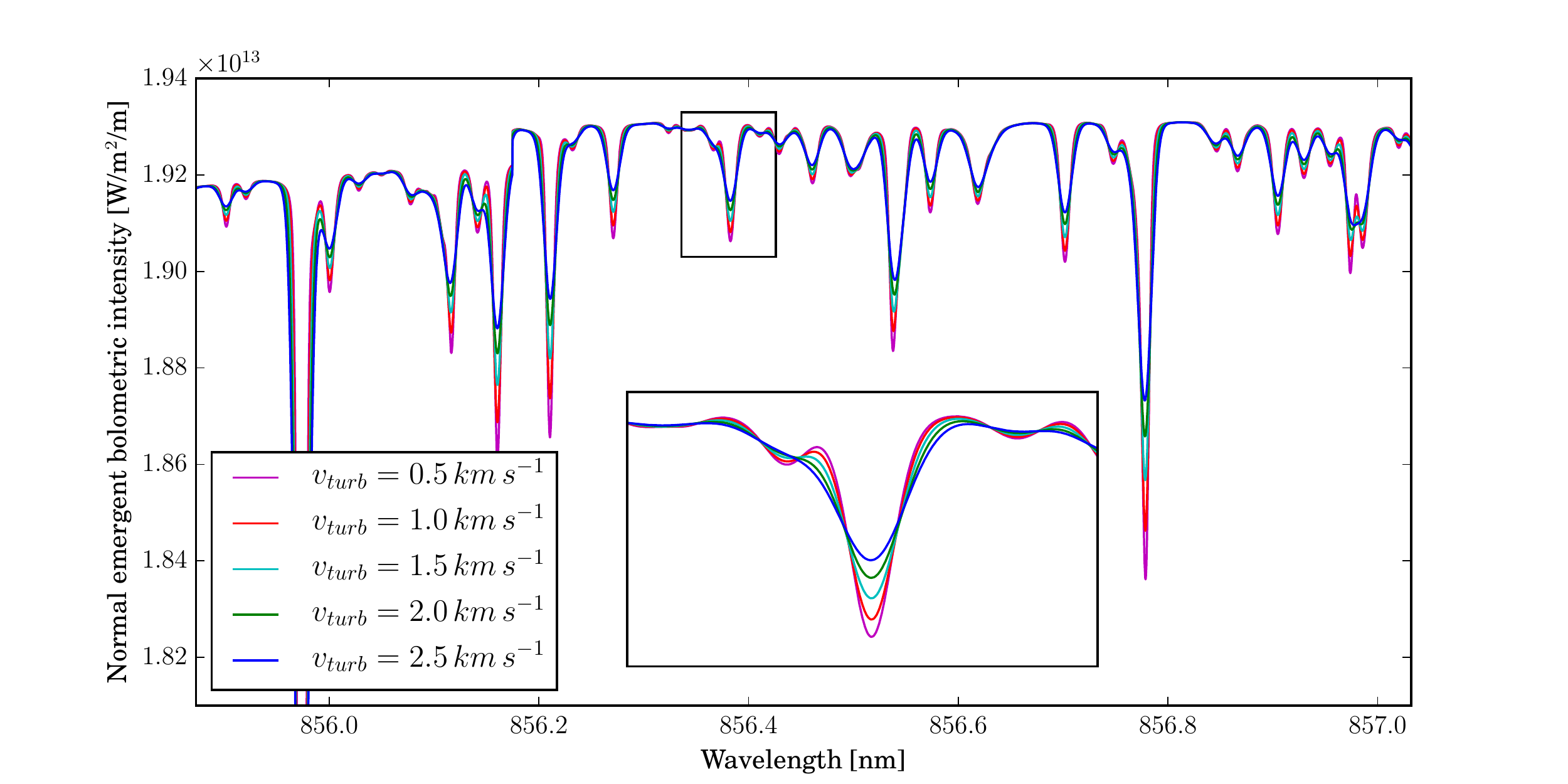}
\end{center}
\caption{
\label{fig:microturbulence}
The effect of microturbulence on the SED. The spectrum is synthesized for the narrow infrared region, at $\Teff=6000$\,K, $\logg = 4.5$ and $\met = 0.0$ using \citet{castelli2004} model atmospheres and the {\sc spectrum} synthesizer \citep{gray1994}. SEDs with $\vturb = 0.5\,\kms$, $1.0\,\kms$, $1.5\,\kms$, $2.0\,\kms$ and $2.5\,\kms$ are plotted. Spectral lines with larger $\vturb$ are shallower and wider, but the integrated passband intensity is equal to within $0.1\%$. Thus, the effect of $\vturb$ on passband intensities is negligible.
}
\end{figure}

Microturbulent velocity is generally an important parameter, but for passband intensities it does not play an appreciable role. This is crucial because it reduces the dimensionality of the interpolation table we need to prepare for calculating intensities. The lack of sensitivity can be understood by considering the effect of microturbulent velocity of spectral lines: the photons that originate at a certain convection cell are emitted from the line core region, but because the mean free path of a photon is larger than the typical size of the convection cell, the photon is absorbed closer to the line wing since the absorption cell is Doppler-shifted w.r.t.~the originating cell. Thus, increasing values of $\vturb$ cause the lines to be shallower and broader, but the equivalent width remains roughly the same (cf.~Fig.~\ref{fig:microturbulence}). This means that the integral does not change appreciably, and the only effect $\vturb$ might have is at the edges of the passband transmission functions. Yet even there the effects are minimal since the transmission efficiency of a typical passband drops gradually.

\subsubsection{Passband Luminosity}\label{sec:pblum}

Passband intensities are computed (and interpolated) in absolute units, but actual data are rarely provided in absolute units. \emph{Passband luminosity} of each star is used to scale the intensities into relative units. The observed luminosity of a star is determined by integrating over all normal emergent passband intensities $I$ multiplied by the integral of the limb darkening model $\mathcal D_\mathrm{int}$ to account for the intensity to flux conversion:
$$ \mathcal L = \int_{\partial V} I(\mu=1; \Teff, \logg, \met) \mathcal D_\mathrm{int}(\Teff, \logg, \met) dA $$
Rescaled intensities are then computed by scaling the observed luminosity $\mathcal L$ to match the prescribed passband luminosity a given passband.

Passband luminosities for each component in a binary can be either coupled or decoupled.  When coupled, the luminosity is provided for one star, and the other star is scaled using the same determined factor. When decoupled, each star is scaled independently with separately provided passband luminosities.

\subsubsection{Limb Darkening}\label{sec:limb_darkening}

Stellar photospheres are opaque to a certain optical depth, so the intensity received along the line of sight to the observer depends on the angle between the normal and the line of sight. If we look parallel to the normal, we see the deepest, hottest layer of the photosphere; this is the normal emergent passband intensity $I_\mathrm{pb}$ given by Eq.~(\ref{eq:passband_intensity}). As the angle increases, we see the progressively shallower, cooler layers, which results in intensity reduction; this is the limb darkened passband intensity $I_\mu$, where $\mu = \cos \theta$ is the measure of the emergent angle. This intensity variation with the distance from the center of the disk depends on geometry, but also on all other complex processes included in the atmosphere models that describe intensity as a function of emergent angle. This is another time-consuming operation, since each local circumstance requires an integration that depends on the angle, i.e.:
\begin{equation} \label{eq:ld_intensity}
I_\mu (\Teff, \logg, \met, \vturb, \mu, \dots) = \int_\lambda \mathcal S_\mu (\lambda; \Teff, \logg, \met, \vturb, \mu, \dots) \mathcal P(\lambda) d\lambda,
\end{equation}
where $S_\mu \equiv dI_\mu(\lambda)/d\lambda$ is the \emph{non}-normal SED function. Traditionally, the functional form of $I_\mu$ is approximated by 1-, 2- or 4-parameter models:

\begin{tabular}{ll}
$\frac{I}{I_0} = 1 - x_\lambda (1-\mu)$ & linear model \\
$\frac{I}{I_0} = 1 - x_\lambda (1-\mu) - y_\lambda \mu \log_{10} \mu$ & logarithmic model \\
$\frac{I}{I_0} = 1 - x_\lambda (1-\mu) - y_\lambda (1-\sqrt{\mu})$ & square  model \\
$\frac{I}{I_0} = 1 - x_\lambda (1-\mu) - y_\lambda (1-\mu)^2$ & quadratic model \\
$\frac{I}{I_0} = 1 - c_{1,\lambda} (1 - \mu^{\frac{1}{2}}) - c_{2,\lambda} (1-\mu) -c_{3,\lambda}(1-\mu^{\frac{3}{2}}) - c_{4,\lambda} (1-\mu^2)$ & power model \\
\end{tabular}

\bigskip

\begin{figure}[t!]
\begin{center}
\includegraphics[width=\textwidth]{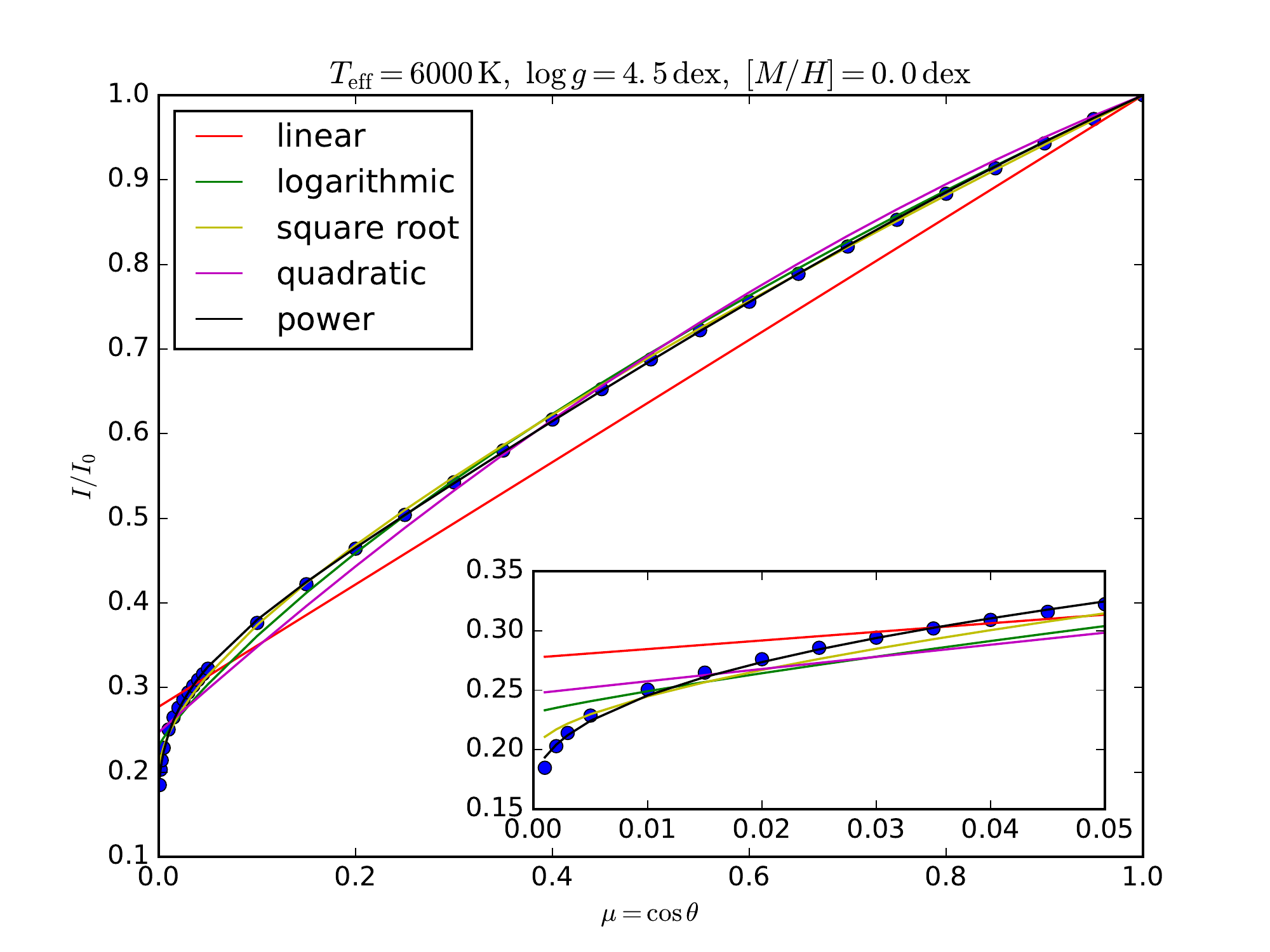}
\end{center}
\caption{
\label{fig:ldfit}
Variation of emergent passband intensity (w.r.t.~normal emergent passband intensity) with emergent angle $\mu \equiv \cos \theta$. The inset is the zoomed in version of the bottom left part of the diagram. Filled circles are integrated intensities (cf.~Eq.~\ref{eq:ld_intensity}) based on \citet{castelli2004} model atmospheres, and lines correspond to the fitted limb darkening models. Systematics that arises from using these low-parametric models can be significant.
}
\end{figure}

Fig.~\ref{fig:ldfit} depicts an example of limb darkening for a solar-like star, along with the fits of the models listed above. Emergent intensities $I_\mu$ (solid circles) are computed using the \citet{castelli2004} models and the SPECTRUM synthesizer \citep{gray1994}. The sequence of 32 $\mu$ values at which we compute emergent intensities is denser towards the limb where the variation is the greatest. Model fits are computed by the unweighted Levenberg-Marquardt method, and the resulting coefficients ($x_\lambda$, $y_\lambda$, $c_{i,\lambda}$) are stored in a table for interpolation purposes.

It is instructive to consider the implications of using these low-parametric models to approximate limb darkening. It is immediately obvious that a linear limb darkening model is inadequate for most practical purposes; two-parameter models are reasonable for angles close to normal but can fail dramatically as $\mu \to 0$. The power model has the smallest residuals, but it also fails as $\mu \to 0$ since it does not drop to $I_\mu/I_0 = 0$ fast enough. This introduces systematic uncertainty to the ingress and egress parts of eclipses, and is particularly dire for planetary transits, where the size of the planet depends critically on the precise treatment of limb darkening. Thus, breaking with the tradition, \phoebe by default uses the precomputed grid of all 32 emergent angles and interpolates in all 4 dimensions ($T_\mathrm{eff}$, $\log g$, [M/H] and $\mu$) to attain $I_\mu$ as accurately as possible. Alternatively, limb darkening models listed above can also be used by passing suitable coefficients ($x_\lambda$, $y_\lambda$, $c_{i,\lambda}$), which remains practical for two reasons: (1) it enables a direct comparison with traditional EB models, and (2) it provides an option to break away from the exclusively model atmosphere-driven results. While the latter provides self-consistency, it also exposes the solution to any systematics that are inherited solely from the choice of the model atmospheres.

\subsubsection{Reflection effect}

A part of the emitted light of each component in a binary system that is directed towards its companion star is reflected off of the companion's surface. This effect needs to be taken into account when we determine \emph{radiosity} $F$, defined as the radiant flux leaving (emitted, reflected and transmitted by) a surface per unit area. For this purpose, we treat the meshes of the bodies as diffusely reflecting and radiative surfaces. The heating of the stars caused by the absorbed part of the light is currently neglected and planned for future extension (Horvat et al., in preparation).

For reflection purposes, we define \emph{intensity} as flux per unit solid angle per unit projected area, \emph{irradiance} as radiant flux received per unit area, and \emph{exitance} as radiant flux emitted per unit area at each surface point. The irradiance can be expressed as a function of the exitance using Lambert's cosine law or a limb-darkening law. Thus far, the reflection effect has been approximated by \citet{wilson1990}'s model, which assumes that the total radiosity of a given surface is composed of the intrinsic emittance and the reflected irradiance as functions of diffusely radiated radiosity (given by the intrinsic intensities of the surface points) following a limb darkening law. There is no strong physical justification for this treatment of reflected light because the reflected and intrinsic light are not directly related and have to be dealt with separately. In order to overcome this issue, we have developed a new reflection model in which the irradiance is given as a function of two terms: the intrinsic emittance following the limb darkening law and the irradiance diffused by the Lambertian cosine law. In this approach, the bodies act as limb darkened radiators of intrinsic light and ideal Lambertian diffusive radiators of incoming light. A full description of the new model is given in Appendix \ref{appendix:reflection_models}.

\begin{figure}[t!]
\begin{center}
\includegraphics[width=\textwidth]{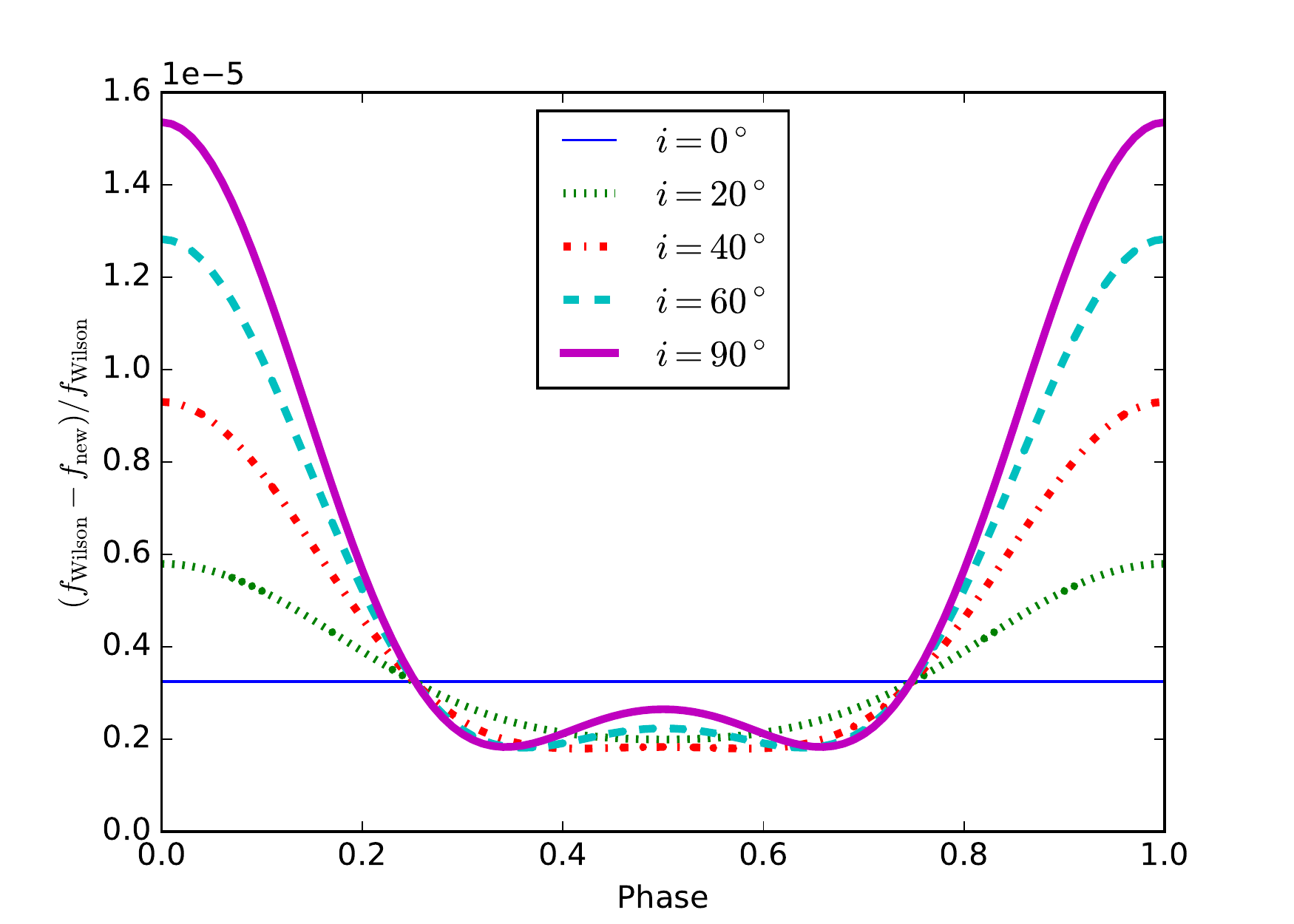} \\
\end{center}
\caption{
\label{fig:reflcomp}
\rev{Relative flux difference between the \citet{wilson1990} treatment of reflection and Lambertian reflection for an A0-K0 system in a 1-day circular orbit. The differences are plotted for several values of inclination, where $i=0^\circ$ is face-on and $i=90^\circ$ is edge-on orientation. Eclipses are \emph{not} included in the computation. The baseline difference at $i=0$ is positive, implying that the contribution to the total flux by reprocessed flux computed by Wilson's scheme is larger than by scattered Lambertian flux computed by our method. This is a consequence of the limb darkening model assumed by Wilson's scheme.}
}
\end{figure}

\phoebe supports both Wilson's model and the Lambertian model presented here. Although the quantitative differences are typically small \rev{(of the order of $10^{-5}$; cf.~Fig.~\ref{fig:reflcomp})}, the new approach is a step towards a complete reflection/heating/redistribution balance equation \rev{(Horvat et al., in preparation)}, it is conceptually more correct and can be applied to custom atmosphere intensities.

\subsubsection{Doppler boosting}

The amount of light that reaches the observer is affected by the kinematic properties of the radiating body. There are three important velocity-induced changes to the received flux: (1) Doppler shift of the spectrum causes the passband-weighted integral  to change (Eqs.~\ref{eq:passband_intensity} and \ref{eq:passband_intensity_photon}); (2) Doppler shift of the frequency causes arrival rate of the photons to change because of time dilation; and (3) relativistic beaming, where radiation is no longer isotropic but becomes direction-dependent because of light aberration \citep[Chapter 4.8]{rybicki1979}. \citet{loeb2003} showed that the boosting signal in exoplanet-hosting stars is larger than variability due to reflected light from the planet; \citet{zucker2007} did a similar study for binary stars and showed that ellipsoidal variability, reflection and boosting are the same order-of-magnitude effects in binaries with a $\sim$10-d orbital period, while boosting dominates in binaries with a $\sim$100-d orbital period. Studies by \citet{vankerkwijk2010}, \citet{bloemen2011} and \citet{bloemen2012} apply this correction to {\sl Kepler} objects KOI-74, KOI-81 and KPD 1946+4340 and demonstrate that the contributions of boosting are indeed crucial to model the light curves correctly.

\begin{figure}[t!]
\begin{center}
\includegraphics[width=\textwidth]{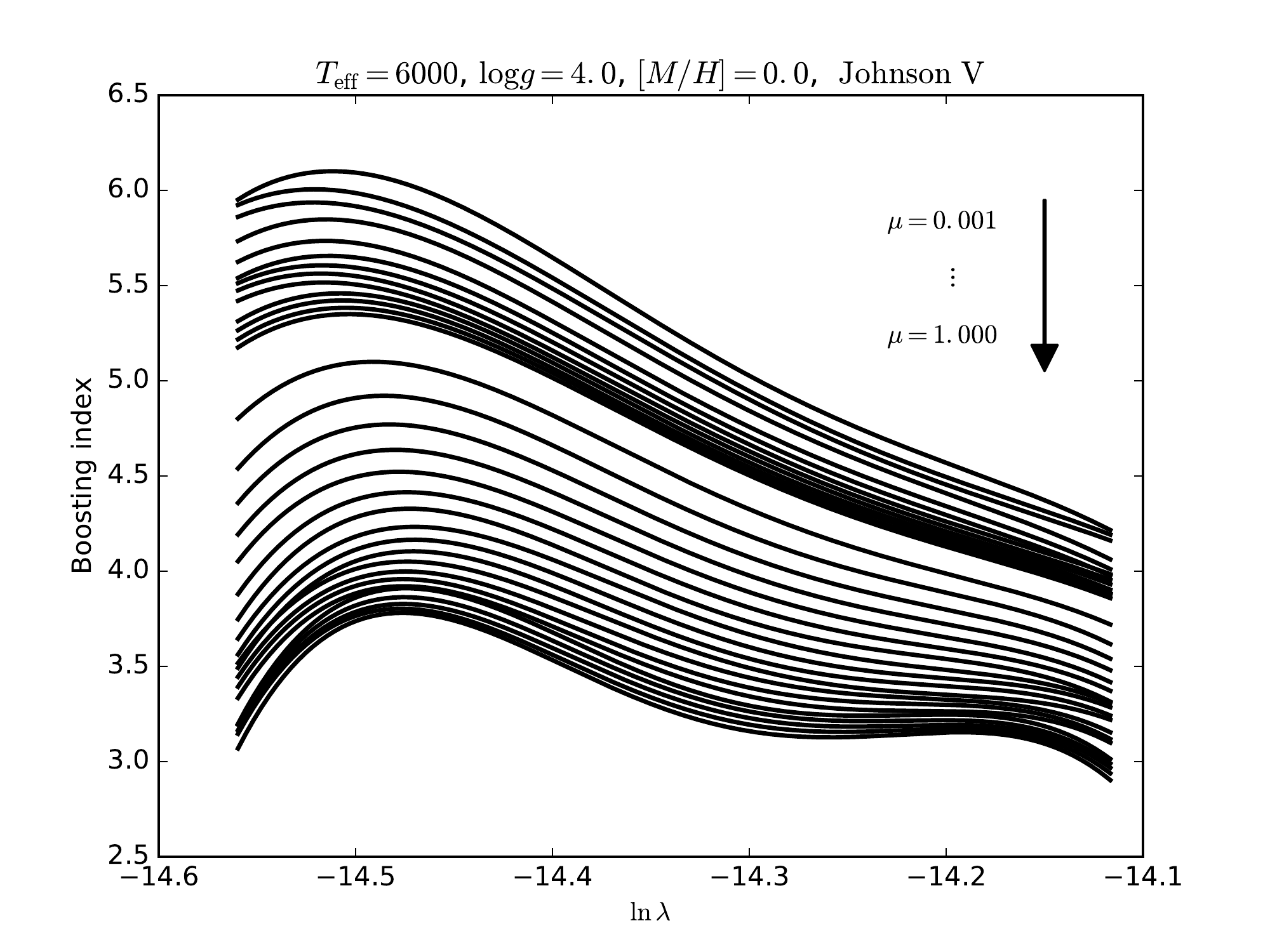}
\end{center}
\caption{
\label{fig:boosting_index}
Boosting index $B(\lambda)$ for the Johnson V broadband filter. The spectra correspond to $\Teff = 6000$\,K, $\logg = 4.0$, [M/H]$=0.0$ and $\mu$ runs over 33 values ranging from 0.001 to 1.0. The dependence of the boosting index on wavelength is obvious, but at this time we approximate it with a weighted average across the passband (cf.~Eq.~(\ref{eq:passband_index})). The increased curviness at larger $\mu$ is due to the larger variations and larger fluxes as we look towards the center of the stellar disk.
}
\end{figure}

The combined Doppler boosting signal can be written as:
\begin{equation}
  I_\lambda = I_{0,\lambda} \left( 1 - B(\lambda) \frac {v_r}{c} \right),
\end{equation}
where $I_\lambda$ is the boosted passband intensity, $I_{0,\lambda}$ is the initial passband intensity, $v_r$ is the radial velocity, $c$ is the speed of light and $B(\lambda)$ is the boosting index:
\begin{equation} \label{eq:boosting_index}
  B(\lambda) = 5 + \alpha \equiv 5 + \frac{d \ln I}{d \ln \lambda},
\end{equation}
where $\alpha \equiv d(\ln I)/d(\ln \lambda)$ is the spectral index, and 5 comes from the Lorenz invariance of $I \lambda^5$ \citep{loeb2003}. Fig.~\ref{fig:boosting_index} depicts the boosting index for the broad Johnson V passband, for a series of $\Teff = 6000$\,K, $\logg = 4.0$, [M/H]$=0.0$ spectra where $\mu$ ranges from 0.001 to 1.0. The dependence of the boosting index on wavelength is notable, but in the simplified approach we approximate the monochromatic boosting factor $B(\lambda)$ with the passband-averaged value:
\begin{equation} \label{eq:passband_index}
  B_\mathrm{pb} = \frac{\int_\lambda \mathcal P(\lambda) \mathcal S(\lambda) B(\lambda) d\lambda}{\int_\lambda \mathcal P(\lambda) \mathcal S(\lambda) d\lambda}
\end{equation}
(or its photon-weighted counterpart per Eq.~\ref{eq:passband_intensity_photon}). We do this by fitting a Legendre polynomial of the 5th order to the $\mathrm{ln}\,I_\mu (\mathrm{ln}\,\lambda$ data; the order is determined by evaluating the rank of the coefficient matrix in the least squares fit and seeing where the values become susceptible to numerical noise. The series was fit iteratively as data points that lie below the $-1\sigma$ threshold were discarded by sigma-clipping the dataset. The iteration stopped when no further points are removed. Another benefit of Legendre polynomials is their analytical derivative; we use this analytic form do derive the average boosting index for every model atmosphere. Fig.~\ref{fig:boosting_index} also shows that the dependence of the boosting index on $\mu$ is significant, so the traditional approach of using integrated flux SEDs to estimate $B_\mathrm{pb}$ is less precise than the treatment presented here.

\begin{figure}[t!]
\begin{center}
\includegraphics[width=\textwidth]{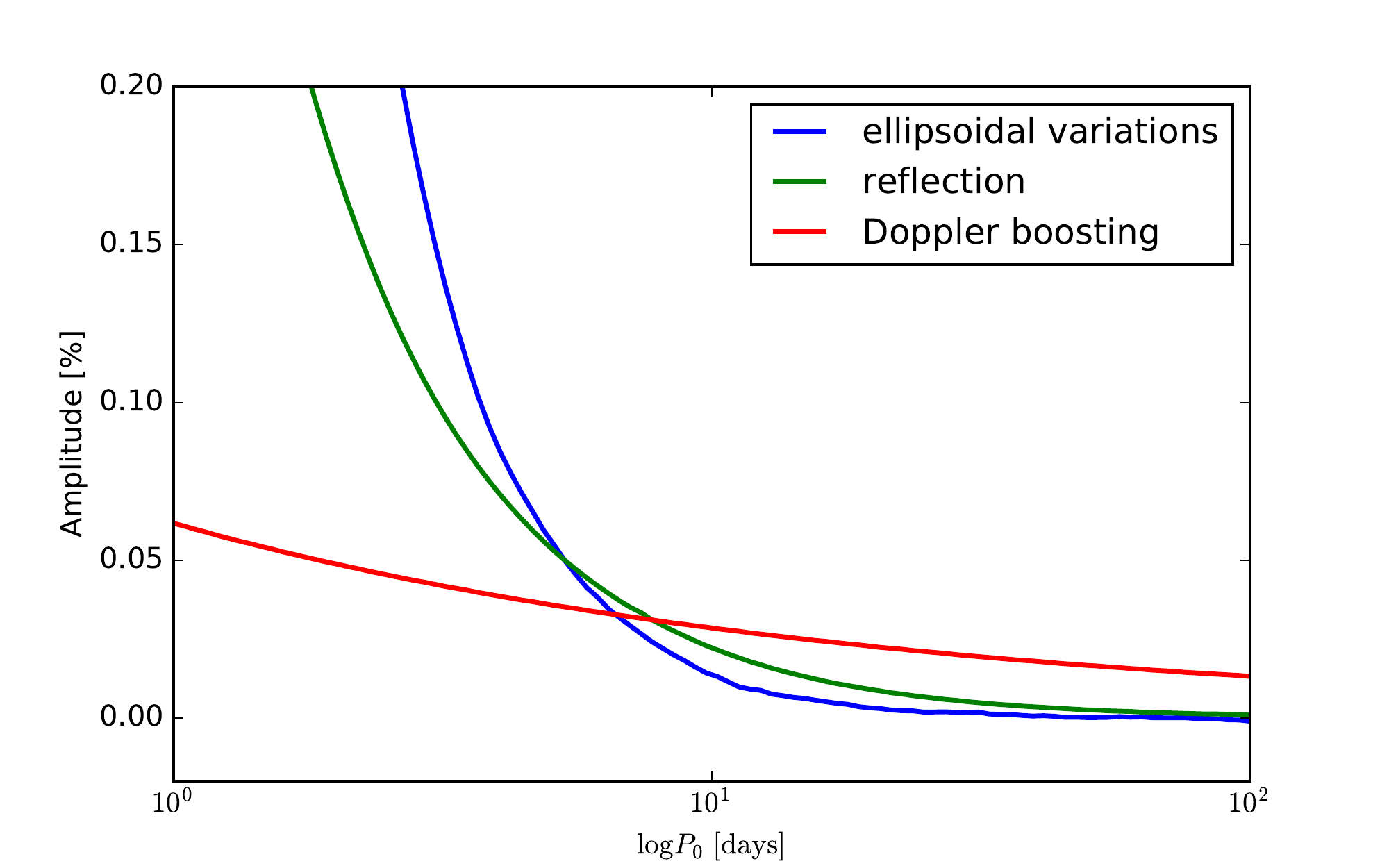} \\
\end{center}
\caption{
\label{fig:amplitude_comparison}
Amplitudes of ellipsoidal variation (blue), reflection (green) and boosting (red) for an A0-K0 main sequence pair \rev{in the Johnson V passband} as a function of orbital period. The semi-major axis is constrained to conserve the masses of components. For this particular system, ellipsoidal variations dominate the short-period end, reflection takes over at $\sim$5.3 days and boosting takes over at $\sim$7.7 days.
}
\end{figure}

\begin{figure}[t!]
\begin{center}
\includegraphics[width=\textwidth]{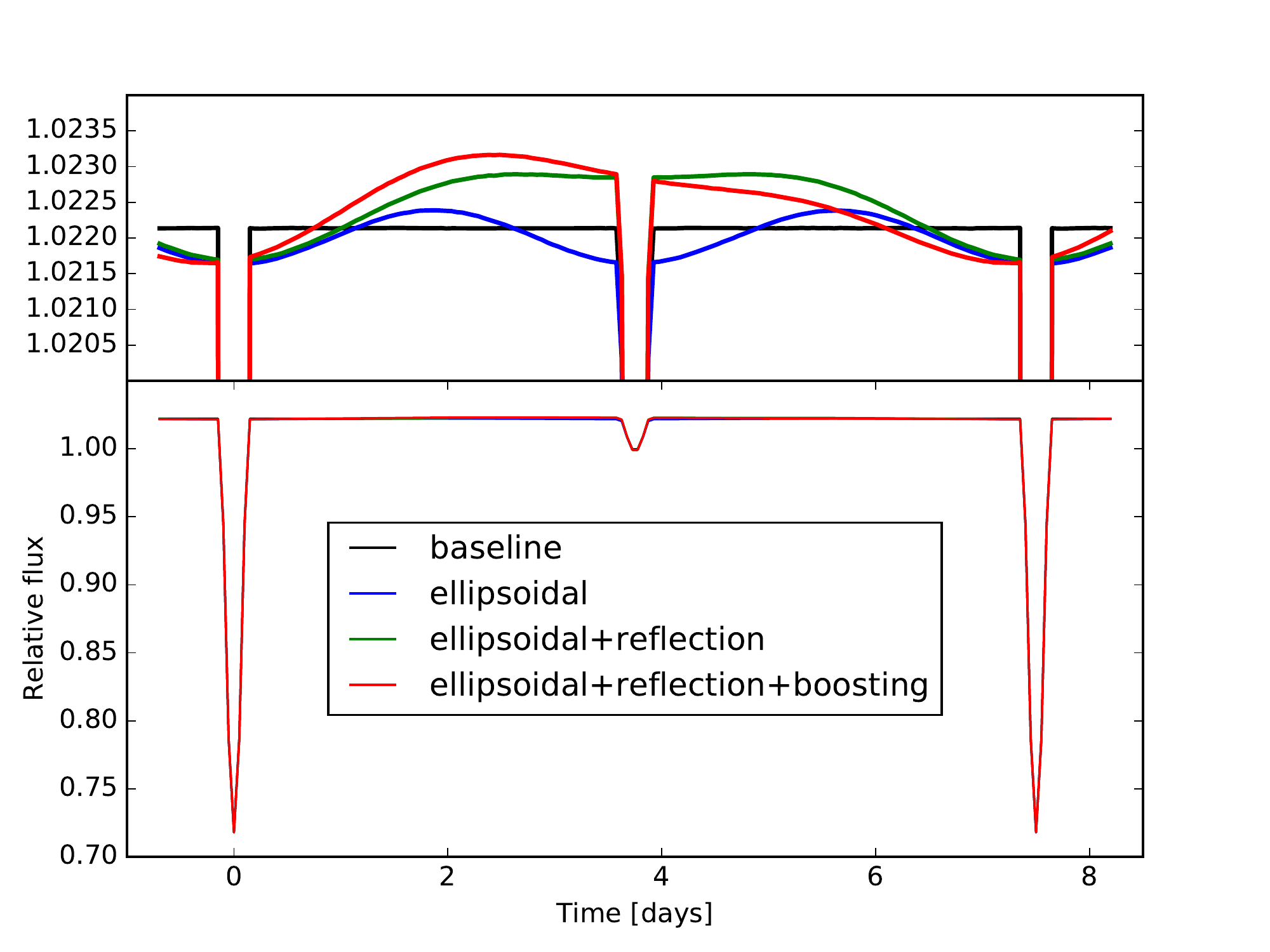} \\
\end{center}
\caption{
\label{fig:boosting_comparison}
Light curve of an A0-K0 main sequence pair in a 7.5-day orbit. The top panel is the zoomed in version of the bottom panel. Four models are being plotted: the baseline model (black), which assumes spherical stars; the ellipsoidal variation model (blue), which includes surface deformation due to tides and rotation; the ellipsoidal variation model with reflection (green), which introduces reflection; and the complete model (red), which accounts for Doppler boosting on top of other effects. While the effects are small in amplitude as evident from the bottom panel, high precision (mmag or better) photometry will routinely demonstrate these effects.
}
\end{figure}

Boosting is a significant effect, and becomes a dominant effect over ellipsoidal variation and reflection for longer period systems. Fig.~\ref{fig:amplitude_comparison} shows a comparison in the amplitude of ellipsoidal variation (blue), reflection (green) and Doppler boosting (red) as a function of orbital period for an A0-K0 main sequence pair in the Johnson V passband. Ellipsoidal variation and reflection dominate the short period end, with boosting taking over at around 8 days. Fig.~\ref{fig:boosting_comparison} depicts a light curve with no effects computed (black), with only ellipsoidal variations (blue), with ellipsoidal variations and reflection (green), and with ellipsoidal variations, reflection and boosting (red).

\subsection{Local radial velocities}\label{sec:velocities}

Local radial velocities are computed for each vertex in the mesh according to the synchronicity parameter $F$ (rotational velocity), systemic velocity $v_\gamma$, and the instantaneous orbital velocity.  Any velocity caused by the changing shape of the mesh itself (i.e.~due to volume conservation for eccentric orbits discussed in Sec.~\ref{sec:volume_conservation}) are currently neglected.

Optionally, gravitational redshift can be accounted for in the radial velocities (which does not affect the reported $v_z$ values), in which case the stored radial velocities are redshifted due to the mass of the originating surface gravity of the star:
\begin{equation} \label{eq:rv_grav}
    \mathrm{RV} = v_z + GMc/R_\mathrm{pole}
\end{equation}

\subsection{Eclipse and horizon computation}\label{eclipse detection}

In order to determine the total flux received from the system, intensities must be integrated over the visible surface elements at any given time. \phoebe determines the visibility of each mesh triangle though horizon and eclipse computation, which is performed once the meshes of all bodies are placed in orbit and their physical quantities computed for each vertex.

Eclipse computation is done on the offset triangulated mesh (cf.~the discussion in the beginning of Section \ref{s:local_quantities}). Fig.~\ref{fig:horizon} depicts the projected shadow of a marching mesh compared to the theoretical horizon (see Appendix \ref{appendix:horizon}) and demonstrates that the determined horizon of each star agrees well with the theoretical expectation. In contrast, the Fourier-based eclipse detection implemented in PHOEBE 1 (and WD) determines the horizon of each star by fitting a Fourier series to the coordinates of surface elements closest to the horizon \citep{wilson1993}. In consequence, this approach will always underestimate the size of the shadow with respect to the analytic horizon. This underestimation converges with an increase in the number of surface elements, but very slowly.  Due to the offsetting of surface elements to obtain the correct numerical surface area, \phoebe does not underestimate the horizon even for a very coarse mesh and increasing the number of elements in \phoebe only makes the horizon smoother.

\begin{figure}[t!]
\begin{center}
\includegraphics[width=\textwidth]{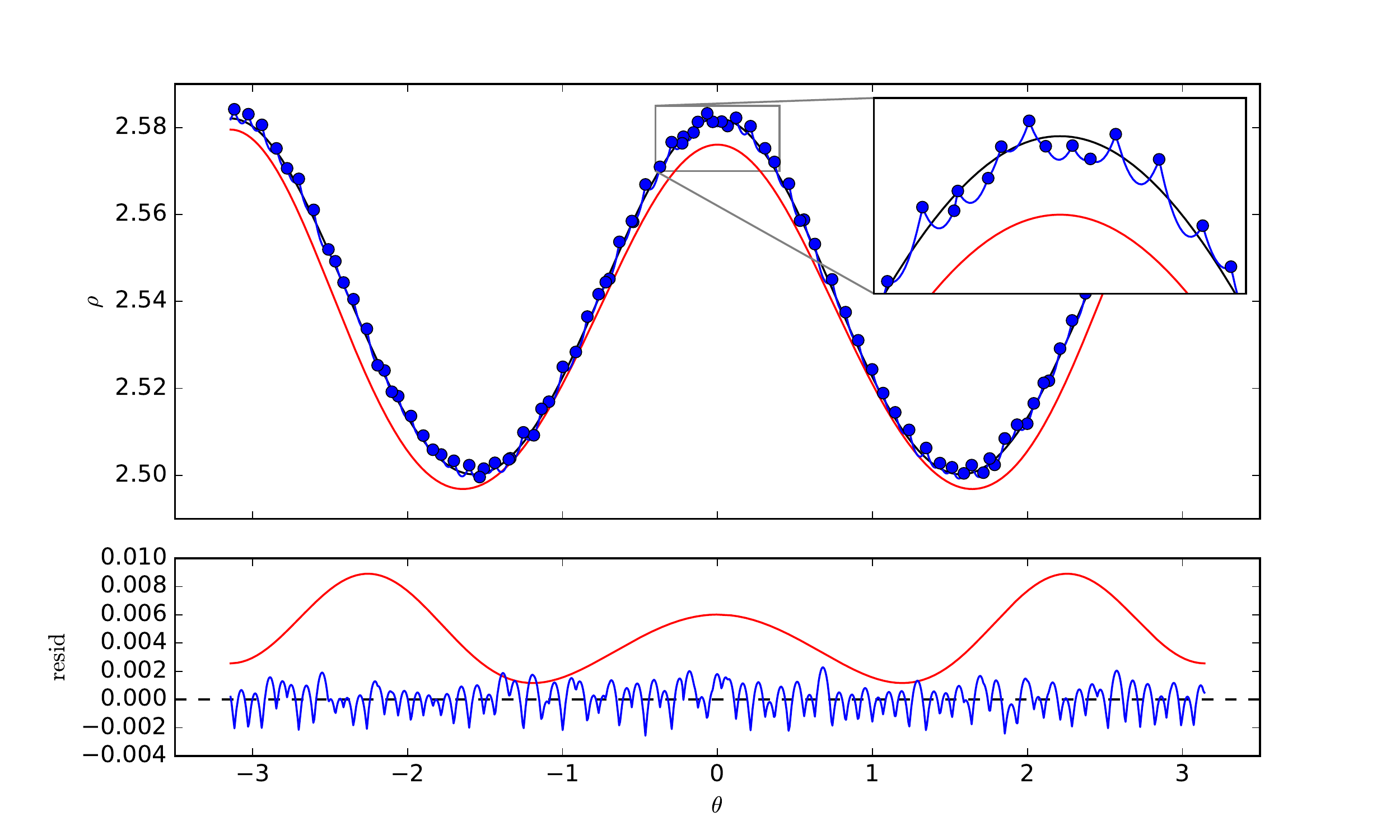}
\end{center}
\caption{
\label{fig:horizon}
Horizon of a Roche star as determined by PHOEBE and WD in polar coordinates.  The black curve represents the analytic horizon (described in Appendix \ref{appendix:horizon}), the red curve represents the Fourier fit to the horizon used in WD, the blue points represent surface elements of the offset PHOEBE mesh that are on the horizon, and the blue line represents the polygonal shadow that these points span. The lower panel shows the residuals between the Fourier-based (red) and offset triangulation-based (blue) horizons with respect to the analytical horizon.
}
\end{figure}

Eclipse detection of triangular meshes in \phoebe is performed algebraically. First, we do back-face culling of triangles, based on the normals of the underlying smooth surface, to obtain a list of potentially visible triangles. Then we utilize Painter's algorithm \citep{hughes2013}, whereby these triangles are ordered w.r.t.~distance from the observer and projected onto the plane-of-sky. Both steps are standard techniques for hidden surface removal in computer graphics. Next, we calculate the visible part of the projected triangles using the clipping algorithm adapted from the  Clipper 2D polygon algebra library\footnote{\url{http://www.angusj.com/delphi/clipper.php}}, which is based on Vatti's method \citep{vatti1992}. From the visible parts of the projected triangles we deduce the ratio of each triangle that is visible as well as the revised centroid of the visible portion by adapting the implementation by Paul Bourke\footnote{\url{http://paulbourke.net/geometry/polygonmesh/}}. 

\begin{figure}[t!]
\begin{center}
\includegraphics[width=0.8\textwidth]{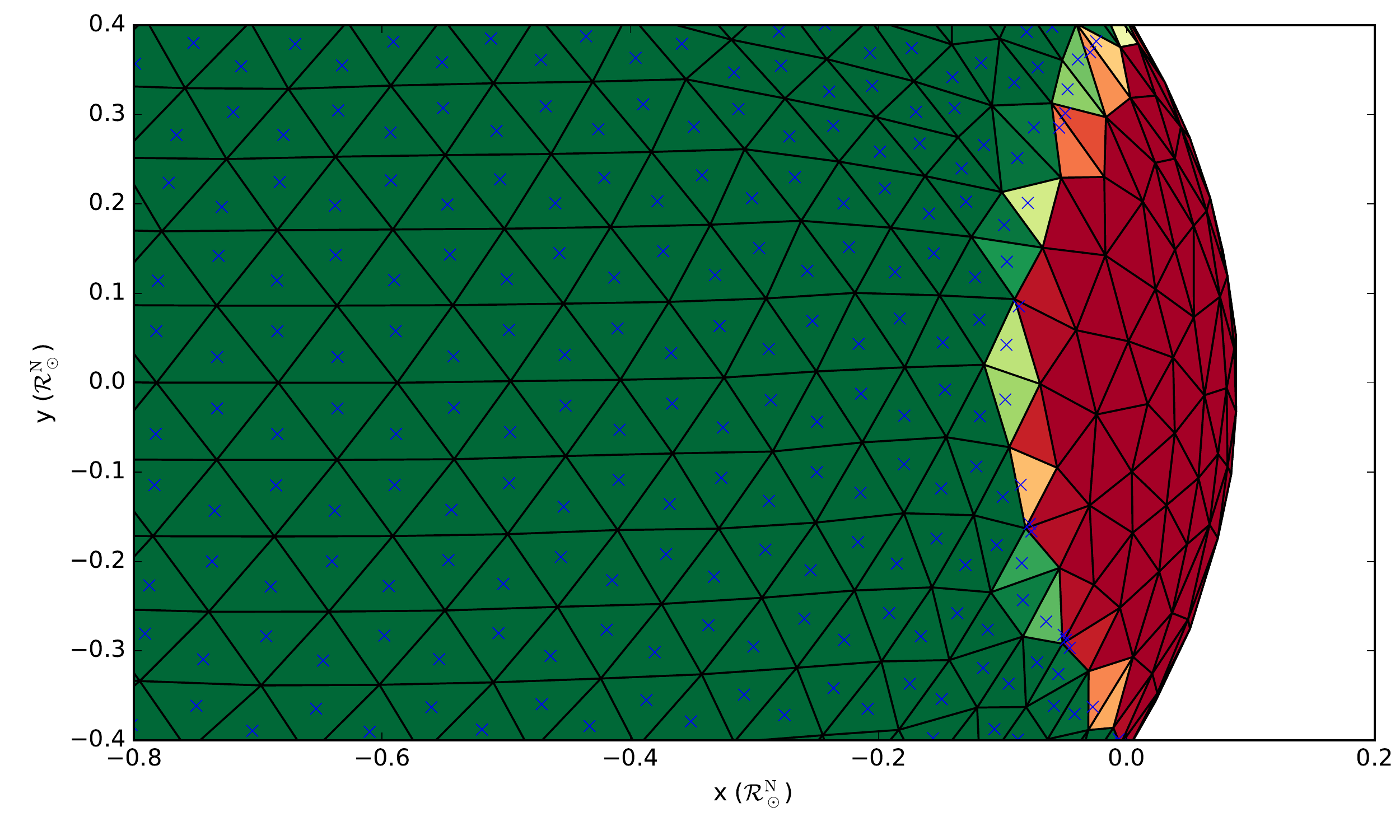}
\end{center}
\caption{
\label{fig:centroids}
A segment of a partially-eclipsed mesh.  The colors of each triangle represent the ratio of the triangle area that is currently visible, with green being fully visible and red being fully eclipsed. The blue crosses represent the centroid of the visible portion of the triangle.  For fully visible (green) triangles, the centroids appear at geometric centers of the triangles whereas the partially visible (orange and yellow) triangles show the centroid moving away from the eclipsed region. These centroids and ratios are used to integrate over the visible surface of the star(s).
}
\end{figure}

Any given \emph{observable} quantity of each triangle is then determined by arithmetic mean of that quantity at each vertex to determine the value at the centroid, under the assumption that these properties vary linearly across the local surface element (cf.~Fig.~\ref{fig:centroids}). Its observable area is then determined based on the provided ratio. This method allows for the accurate approximation of ingress/egress without the requirement of a fine mesh that is more computationally expensive.

\subsubsection{Flux computation}

The actual observable in light curves is flux. Flux is computed by integrating scaled, limb-darkened, projected intensities along the line of sight over the visible surface of the star. Any extraneous light (passed as a parameter) is then added to this value to obtain the observable flux. The process is then repeated for each time-stamp, where local quantities are updated if the mesh has changed, and re-integrated to obtain a new flux.

\begin{figure}[t!]
\begin{center}
\includegraphics[width=\textwidth]{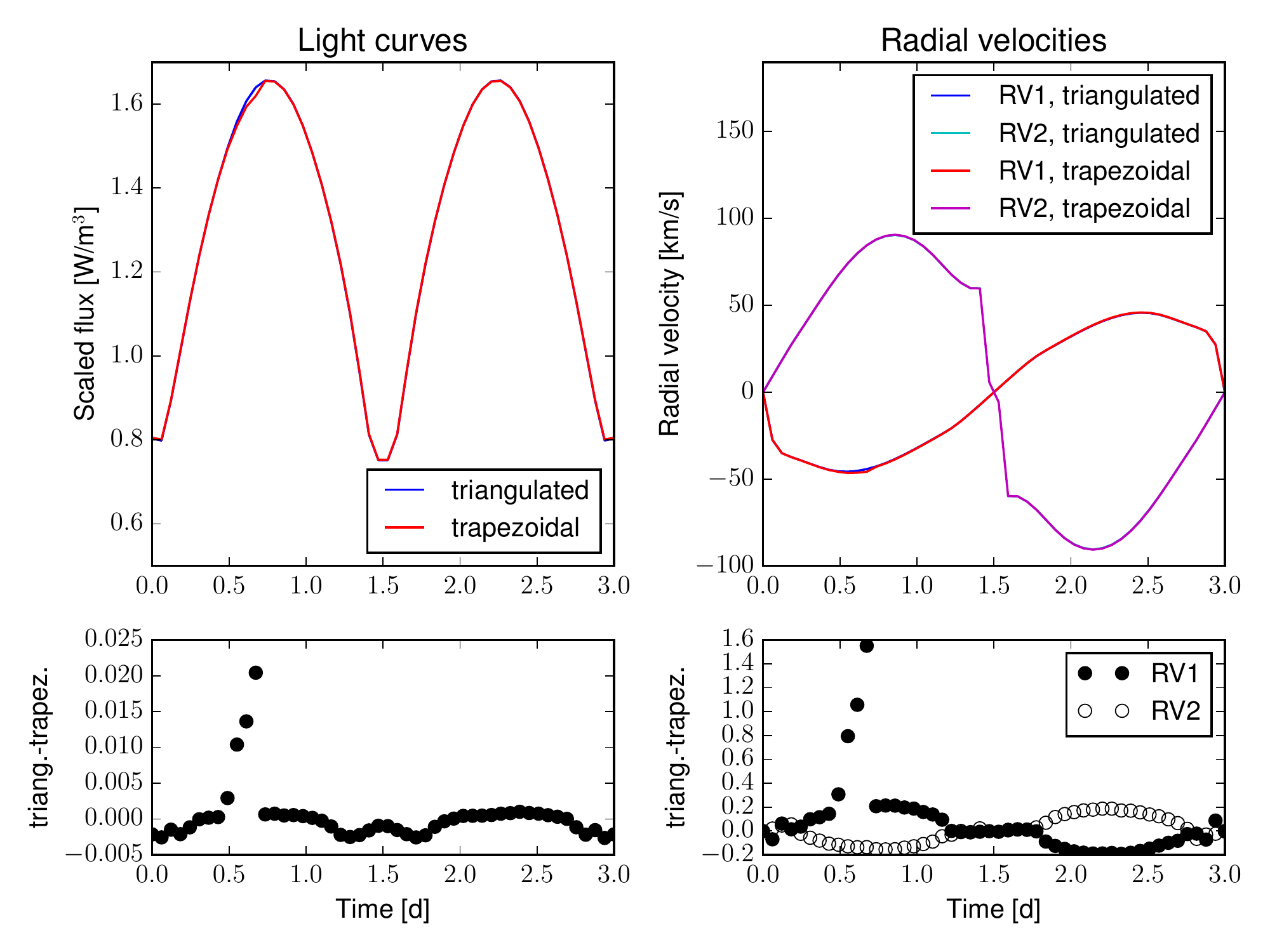}
\caption{\rev{Light curves and radial velocities of the triangulated and trapezoidal meshes of an contact system with mass ratio $q = 0.5$, fillout factor $FF = 0.9$, equal surface temperatues $T_1 = T_2 = 6000\;\mathrm{K}$ and inclination $i=90^0$.}}
\label{fig:lcs_rvs_comparison}
\end{center}
\end{figure}

\rev{A comparison between the resulting light curves of triangulated and trapezoidal meshes is depicted on the left panels of Fig.~\ref{fig:lcs_rvs_comparison}. As a test case we use a contact system due to its highest level of distortion. The resulting light curves agree to $\sim0.1\%$, with the exception of several discrepant points for which the trapezoidal mesh gives fluxes lower than excepted. This is due to the diverging centers which leave gaps in the mesh, as mentioned in Section~\ref{s:discretization}.}

\begin{figure}[t!]
\begin{center}
\includegraphics[width=\textwidth]{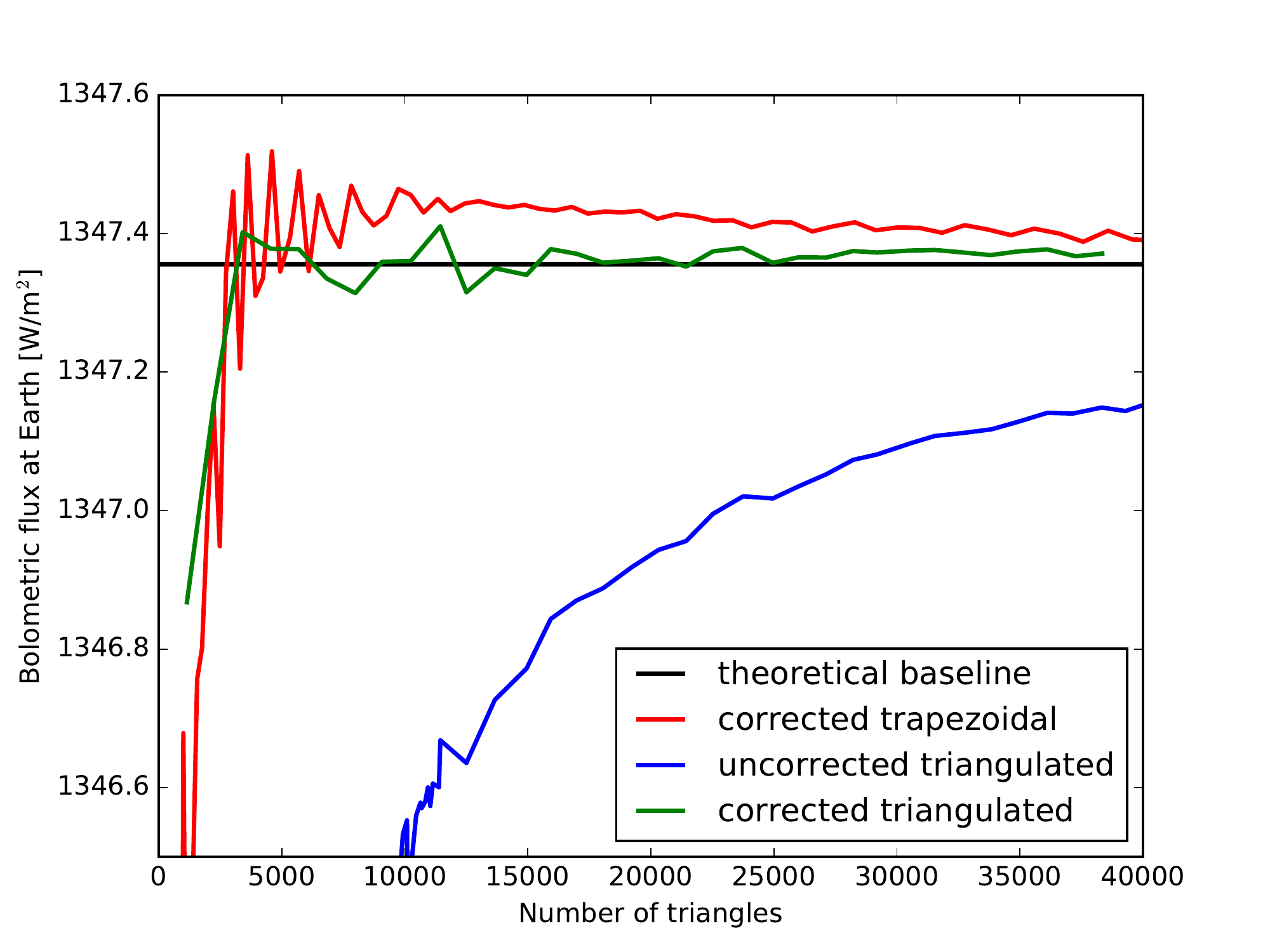} \\
\caption{
\label{fig:flux_comparison}
Comparison of fluxes computed analytically (black line) and for different meshes: trapezoids with analytical areas (red line), triangulation (blue line) and offset triangulation (green line) as a function of mesh size. Corrected (offset) triangulated meshes are consistent with the benchmark and reach it for a smaller number of surface elements than any other method.
}
\end{center}
\end{figure}

To evaluate the precision of flux computation we consider the Sun--Earth system. This is a convenient test case because it is the best calibrated benchmark we have, and because it pushes model parameters to the extreme operation mode. We limit our test to the blackbody model because it is the only model that we can integrate analytically. Fig.~\ref{fig:flux_comparison} summarizes the results. The \rev{black} line is the \rev{theoretical} baseline computed by analytical integration. Other curves represent the results from meshed models and they depict fluxes as a function of the number of surface elements. Offset triangulation, explained in Appendix \ref{appendix:offsetting}, performs most robustly and it reproduces the analytical value effectively.

\subsubsection{Radial Velocity computation}

\phoebe computes radial velocities (RVs) in two regimes: dynamical and integrated. Dynamical RVs correspond to the line-of-sight component of the orbital velocity vector of each star; since they are analytically computed, they do not exhibit any proximity effects (i.e.~the Rossiter-McLaughlin effect). In order to model these effects, the RVs must also be sampled over mesh of each star; we refer to these as local RVs. Local RVs are assigned to each element by computing the projected velocity (already gravitationally redshifted; cf.~Section \ref{sec:velocities}) at the visible centroid of each triangle, and then integrated over the visible area to the photometrically weighted, integrated RV curve.

\rev{The comparison of integrated RVs of a triangulated and trapezoidal mesh for an contact system is given in Fig.~\ref{fig:lcs_rvs_comparison}.}

\subsection{Finite integration time}

When modeling short period systems, a finite exposure time starts to become relevant. This is especially true for \emph{Kepler} data, where CCDs are read out every 6.54 seconds (6.02\,s live time) and then summed into 29.4244-min Long Cadence bins. This implies that every exposure has an effective ~30-min integration time during which any flux variation will be smeared into a single data point. This bears little significance for objects with periods longer than $\sim$10 days, but it becomes important for short period systems because it convolves the true signal with the exposure time boxcar, resulting in phase smearing. The effect is more pronounced for semi-detached and contact binaries, with orbital periods as short as a few hours. To account for the smoothing of observations due to finite exposure times, PHOEBE can oversample the light-curve and provide an averaged value for any given time.

\section{Conclusions}\label{s:limitations}

This paper presented the first step towards the increased fidelity of computed eclipsing binary models. \rev{Triangulation replaces trapezoidal meshing, which provides a robust surface coverage by near-equilateral, near-isometric surface elements and the ability to discretize any 3-D body irrespective of its shape, whereas trapezoidal meshing relied on close-to-spherical shapes. To attain high mesh precision with small number of triangles, we offset mesh elements so that the discretized volume is identical to the analytical equipotential volume. We introduce a new type of body: rotating stars, which eliminates lengthy Roche computations when they are not necessary for the problem at hand. The treatment of dynamics has been updated by including light travel time delay for all bodies in a system, such that time is measured w.r.t.~the barycenter of the system. Time-to-phase conversion accounts for all temporally changing quantities explicitly. In case of eccentric orbits, we provide an argument of why volume conservation is appropriate and we propose a test based on the eccentric ellipsoidal variable systems known as heartbeat stars. Passband intensities are based on the \citet{castelli2004} model atmospheres and values for both energy-weighted and photon-weighted intensities are stored in lookup tables for fast operation. The old treatment of limb darkening caused severe artifacts near the limb of the star because of undersampling; this has been mitigated by providing the option to use actual intensity values as a function of $\mu$ instead of relying on 1-, 2- or 4-parameter models. Reflection effect has been restated to include Lambertian scattering, which paves the way for future inclusion of heat redistribution across the irradiated surface. Doppler boosting (comprised of Doppler shift of the spectrum, time dilation and relativistic beaming) is now fully taken into account. Local radial velocities are now corrected for the gravitational redshift. Eclipse and horizon computation is significantly improved by replacing the old Fourier-based approach with an algebraic scheme that is as accurate as the mesh itself. Finite integration time acts as a smoothing filter on light curves, which is now supported on the synthetic end as well. Finally, flux computation in absolute units has been tested against the Earth-Sun system and the test demonstrates that the fidelity has indeed increased.}

There are several limitations of \phoebe 2.0 that we aim to address within the ongoing project:

\begin{description}

\item[Integrated flux dependence on the number of surface elements.] \rev{Ideally, the computed flux should not depend on the computational aspects such as the fineness of the mesh, but this is an inherent limitation for all numerical schemes. Flux values converge to the analytical value linearly ($\propto 1/N$ where $N$ is the number of surface elements), which is quite slow and is further impacted by other second-order effects, most notably limb darkening. The exact convergence properties are difficult to assess because of the non-linearity of the parameter space, but tests show that for a moderate number of triangles ($\sim 5000$) the attained relative accuracy is of the order of $10^{-3}$. The impact on \emph{relative} light curve shapes (i.e.~flux ratios) is much smaller because all phase points are affected equivalently, but whenever flux values are sought, their dependence on the mesh size should be kept in mind. We are considering acceleration schemes that model the dependence of integrated flux on the number of surface elements by evaluating it for several mesh sizes and then extrapolating it to $N \to \infty$, but no robust solutions have been worked out yet.}

\item[Contact binaries.] The trapezoidal model to modeling contact binaries is likely suitable only for systems in thermal equilibrium, since it introduces unphysical discontinuities in the neck area of systems with unequal temperatures (cf.~Section \ref{s:local_quantities}). Even for systems in thermal equilibrium, the atmosphere tables used to derive the normal emergent intensities are computed assuming single, spherical stars, which introduces additional inconsistencies in the overall treatment of contact systems. We are currently working on the implementation of a more feasible model of contact binaries by studying the radiative transfer in the common envelope based on the structure of the whole star. The structure and radiative transfer computations are performed independently which allows for the exploitation of a wide range of hydro-thermodynamical models and the testing of the surface intensity distribution they give rise to. The full description of this novel model will be the topic of \rev{the paper by} Kochoska et al., in preparation.

\item[Misaligned binaries.] \phoebe is currently based on the Roche geometry, which assumes perfect alignment between the orbital and the rotational axes and handles each component's tidal distortion as due to the companion's point mass. However, careful ground-based studies have been able to discern that the components of close binaries in a number of systems show misaligned rotational and orbital axes \citep{albrecht2011}. Observationally this is most easily discerned via the Rossiter-McLaughlin effect. The misalignment has also been found for a number of ''hot Jupiters`` \citep{hebrard2008}. Consequently, both EBs and transiting exoplanets bring into question the initial conditions for the formation of both binaries and planetary systems. The generalized Roche potential for binary systems where the stellar rotation is not aligned with the orbital revolution is fundamentally different to the currently implemented aligned potential. The modifications to the potential have been derived by \citet{limber1963}, \citet{kruszewski1967} and \citet{kopal1978}. The properties of the critical equipotential lobe and Lagrangian points for circular orbits have been studied in detail by \citet{avni1982}. We are implementing a fully numerical determination of equipotential surfaces in misaligned binaries. These determine the shapes and radiative properties of components in binary stars and will be presented by Horvat et al., in preparation.

\item[Graphical User Interface.] It might come as a surprise that the current version of \phoebe does not come with a standalone graphical user interface. We are working on a web-based user interface that is substantially more flexible than the original interface of \phoebe 1. Its main characteristic is the combination of an interactive shell and a graphical interface that can be either run locally or remotely. The interface will be discussed by Conroy et al., in preparation.

\item[Fitting.] The absence of the fitting interface is by no means an oversight. Deterministic minimizing programs used in the past, most notably Differential Corrections (DC; \citealt{wilson1971}), Nelder \& Mead's Simplex method (NMS; \citealt{prsa2005b}) and Powell's direction set method \citep{prsa2007}, are not suited for the high fidelity demand of modern data, and do not lend themselves to automation that is necessary to address the firehose of data coming our way from surveys such as Gaia or LSST. We consider \phoebe models as \emph{likelihood functions} that, accompanied with an appropriate noise model, should feed probabilistic samplers, such as the Markov Chain Monte Carlo methods. The High Perfomance Computing (HPC) wrapper for \phoebe is available on \phoebe website and should be preferred to deterministic minimizers. The rationale for fitting will be discussed by Pr\v sa et al., in preparation.

\item[Computational time cost.] The computational infrastructure of \phoebe is implemented in the low-level C language for speed, and the interface part is written in the high-level python language that is inherently slow. A number of causal computations are still linked through python, which causes slowdown of the execution time. Yet the dominant source of slowdown is increased model fidelity. Shortcuts taken before are no longer in effect and, even though effects can be turned off by the user, the impact on the overall runtime is significant. That is why the preferred mode of deployment of \phoebe is on HPC clusters.

\end{description}

In conclusion, we respectfully invite the community to provide us with any feedback, criticism and suggested improvements to the model, and to help us critically evaluate model robustness in different operational regimes. \rev{We remind that \phoebe is released as open source under the General Public License, is free and will always remain free, so anyone interested to join in our efforts to provide a robust, general modeling code to the community is welcome to join us.}

\acknowledgements

The development and implementation of PHOEBE has been supported by the NSF AAG grant \#1517474; it also received partial funding from the European Community's Seventh Framework Programme FP7-SPACE-2011-1, project number 312844 (SPACEINN). K.~E.~Conroy is supported under a NASA NESSF Fellowship \#NNX15AR87H. The authors express sincere gratitude to R.~E.~Wilson, W.~van Hamme and S.~Rucinski for useful discussions. \rev{The anonymous reviewer provided us with further suggestions from the user's point of view, which were immensely helpful and highly appreciated.}

\bibliographystyle{apj}
\bibliography{paper}

\appendix

\section{Curvature-dependent offsetting of mesh surface elements}\label{appendix:offsetting}

The triangulation method computes $n$ vertices of the mesh: ${\bf V} = ({\bf v}_1, \ldots, {\bf v}_n)$. These vertices span $m$ triangles: $ T_j ({\bf r}_{c_{j,1}}, {\bf r}_{c_{j,2}}, {\bf r}_{c_{j,3}})$, $j = 1,\ldots, m$, where $c_{j,i} \in [1,n]$ is an index of the $i$-th vertex spanning the $j$-th triangle. The triplets of indices $(c_{j,1}, c_{j,2}, c_{j,3})$ are ordered in a way that the normal of the triangle, which is proportional to the vector product $ ({\bf v}_{c_{j,2}} - {\bf v}_{c_{j,1}}) \times  ({\bf v}_{c_{j,3}} - {\bf v}_{c_{j,1}})$, points outwards w.r.t.~the equipotential. The indices $c_{j,i}$ are stored in the triangle-vertex connection matrix ${\bf C} = [c_{j,i}]$ of the mesh. \phoebe internally handles meshes stored in the face-vertex format given by the pair $({\bf V}, {\bf C})$.

The mesh can only approximate a smooth isosurface of the potential $\potential$ at some reference value $\potential_0$. This implies that the area of the mesh is different from that of the isosurface and it needs to be corrected. This is achieved by linearly shifting each vertex of the mesh to $ {\bf v}_i' = {\bf v}_i + \alpha_i {\bf n}_i$, $i=1,\ldots, n$, where ${\bf n_i} = \nabla \Phi(r_i)/\|\nabla \Phi(r_i)\| $ is a normal to the isosurface in the $i$-th vertex and $\alpha_i$ is the offset length at that vertex. Finding $\alpha_i$ is an optimization problem done under two constraints: every local surface element needs to approximate the isosurface as closely as possible, and the area of the entire mesh matches the isosurface area exactly: ${\rm A} ({\rm Mesh}({\bf V}' , {\bf C})) = {\rm A}({\cal S})$, where ${\bf V}'= ({\bf v}_1',\ldots, {\bf v}_n')$ is a vector of shifted vertices. 

Qualitatively we expect that shifts $\alpha_i$ have the same sign as the isosurface curvature, and that their value should be larger for larger curvatures since larger curvatures imply larger distances to the neighboring triangles. Since the number of triangles is generally large, any local curvature across a single surface element can be considered constant. Thus, we can optimize $\alpha_i$ on a perfectly spherical mesh and apply those corrections to local surface elements. Sphere triangulation using the marching method and recursive subdivision results in a uniform set of triangles; in the limit of large number of triangles we find that optimal shift for all vertices is equal to $\alpha \sim A_t^2/4R$, where $1/R$ is the curvature of the sphere and $A_t$ is the average area of a single triangle. This result can be generalized to the expected functional form $\alpha_i = t H_i S_i$, where $t$ is the sole scalar parameter for all shifts, $H_i$ is the mean curvature of the underlying isosurface at the vertex approximated from the mesh \citep{taubin1995}, and $S_i$ is 1/3 of the area of triangles sharing the $i$-th vertex. As all shifted vertices ${\bf V}'$ are parametrized with a single parameter $t$, we can easily solve the above equation for $\alpha_i$.

\section{Volume and area of isosurfaces}\label{appendix:volume_surfacearea}

The volume $V$, area $A$ and derivative of the volume w.r.t. the potential $V_{,\potential_0} := \partial V/\partial \potential_0$ of a body within an isosurface ${\cal S} = \{{\bf r} : \potential({\bf r})= \potential_0 \} \>$ are essential ingredients of the robust working of \phoebents. These quantities are computed to arbitrary precision directly from the definition of the Roche potential. 

%
%

The volume and its derivative are essential for the process of volume conservation, in which the reference potential $\potential_0^*$ of an isosurface is computed so that its volume matches the reference value $V^*$. This is done by the Newton-Raphson iteration in the form: 
$$
  \potential_{0,k + 1} = 
  \potential_{0,k} - \frac{V(\Phi_{0,k}) - V^*}{V_{,\potential_0}(\potential_{0,k})}
$$
where we assume convergence if $\lim_{k\to\infty} \potential_{0,k} = \potential_0^*$.

\subsection{Generalized Roche lobes}

We discuss isosurfaces of the generalized potential potential $\potential$ given by Eq.~(\ref{eq:roche_potential}) around the reference value of the potential $\potential_0$.  We express the potential in rescaled cylindrical coordinates:
\begin{align}
  \widetilde\potential(\xi, \sigma, \phi) &=  
  \delta \potential(\delta \xi,  \delta \sqrt{\sigma} \cos\phi, \delta \sqrt{\sigma} \sin\phi) \\
  &= \frac{1}{\sqrt{\xi^2 + \sigma}} + q \left(\frac{1}{\sqrt{\sigma + (\xi-1)^2}} - \xi\right) +  \frac{1}{2} B^2 (\xi^2 + \sigma \cos(\phi)^2)
\end{align}
where $\xi=x/\delta$ is the dimensionless $x$ coordinate, $\sigma=(r/\delta)^2$ is the dimensionless square of the radius and $B= F^2 (1+q) \delta^3$ is an auxiliary constant. Additionally, we introduce the rescaled reference value $\widetilde\potential_0 = \delta \potential_0$ corresponding to the Roche lobe of interest.

We can find the lower and upper bounds of the Roche lobe (denoted by $\xi_0$ and $\xi_1$, respectively) along the $\xi$-axis, by solving the equation
$$
  \widetilde\potential(\xi_i, 0, 0) = \widetilde \potential_0 \qquad   i = 0, 1 \>,
$$
which we do using a combination of analytical estimates and general non-linear equation solvers.

The goal is to compute the following Roche lobe quantities at a reference $\widetilde \potential_0$: 
\begin{itemize}
\item Volume:
$$
  V = 2\delta^3\int_{\xi_0}^{\xi_1} {\rm d} \xi \int_0^{\pi/2} {\rm d} \phi\, \sigma(\xi,\phi)  
$$
\item Derivative of volume w.r.t. rescaled reference value:
$$
  V_{,\widetilde\Phi_0} = 
  2 \delta^3 
  \int_{\xi_0}^{\xi_1} {\rm d} \xi 
  \int_0^{\pi/2} {\rm d} \phi\, 
  [\widetilde\potential_{,\sigma} (\xi, \sigma(\xi,\phi), \phi)]^{-1}\>,
$$
\item Surface area:
$$
  A = 4 \delta^2 \int_{\xi_0}^{\xi_1} {\rm d} \xi \int_0^{\pi/2} {\rm d} \phi\, 
  \sqrt{\sigma(\xi,\phi) + \sigma_{,\phi}(\xi,\sigma(\xi,\phi),\phi)^2/(4\sigma(\xi,\phi)) + \sigma_{,\xi}(\xi,\sigma(\xi,\phi),\phi)^2/4}\>,
$$
where the derivatives of the squared radius w.r.t. to $\xi$ and $\phi$ are given by:
\begin{equation}
  \sigma_{,\xi}(\xi,\sigma,\phi)
  = - \frac{\widetilde\potential_{,\xi}(\xi, \sigma, \phi)}
  {\widetilde\potential_{,\sigma}(\xi, \sigma,\phi)}
  \qquad
   \sigma_{,\phi}(\xi,\sigma,\phi)
  = - \frac{\widetilde\potential_{,\phi}(\xi, \sigma, \phi)}
     {\widetilde\potential_{,\sigma}(\xi, \sigma,\phi)} 
  \label{eq:R_derivatives}
\end{equation}
\end{itemize}

Note that the derivative of the volume w.r.t.~non-rescaled reference value, needed in volume conservation, is simply $V_{,\potential_0} = \delta V_{,\widetilde\potential_0}$. All these quantities have a common functional form and can we written as an integral of some vector function $\bf F$:
\begin{align*}
  {\bf I} &:= (V,V_{,\widetilde\potential_0}, A) \>, \\
         &= \int_{\xi_0}^{\xi_1} {\rm d} t \int_0^{\pi/4} {\rm d} \phi\, 
   {\bf F}(\xi, \sigma(\xi, \phi), \phi) \>.
\end{align*}
We can express $\bf I$ as an integral of its derivative along the $t$ axis:
$$
  {\bf I} = \int_{\xi_0}^{\xi_1} {\rm d} \xi\, \dot {\bf I}(\xi)\quad {\rm and}\quad
  \dot {\bf I}(\xi) = \int_0^{\pi/4} {\rm d} \phi\, {\bf F}(\xi, \sigma(\xi, \phi), \phi)\>
$$
and with the use of a Legendre-Gauss quadrature of order $n$ (see e.g. in \citealt{sirca2012}), we can approximate the integral of $\dot {\bf I}$ over $\phi \in [0,\pi/4]$ as a sum:
\begin{equation}
 \dot {\bf I}(\xi) \approx \sum_{i=1}^n w_i {\bf F}(\xi, \sigma(\xi, \phi_i), \phi_i)\>,
 \label{eq:I_sum}
 \end{equation}
where the weights $w_i$ and angles $\phi_i \in [0, \pi/4]$ are calculated for the chosen quadrature rule and adapted to our case. With the introduction of a squared-radius at specific angles $\sigma_i(\xi) :=  \sigma(\xi, \phi_i)$, Eqs.~(\ref{eq:R_derivatives}) and (\ref{eq:I_sum}) yield a system of $n+3$ ordinary differential equations for $\sigma_i$ and $\bf I$:
\begin{align}
  \dot \sigma_i(\xi) &= \sigma_{,\xi}(\xi, \sigma_i(\xi), \phi_i)\qquad i=1, \ldots,n\>, \\
  \dot {\bf I}(t) &= \sum_{i=1}^n w_i {\bf F}(\xi, \sigma_i(\xi), \phi_i)
\end{align}
We integrate the above system of equations in times $\xi \in [\xi_0,\xi_1]$ with initial conditions $\sigma_i(\xi_0) = 0$ for $i=1,\ldots,m$ and ${\bf I}(\xi_0) = 0$. The Roche lobe quantities are given by ${\bf I}(\xi_1)= (V,V_{,\widetilde\potential_0}, A)$. In \phoebents, we use Legendre-Gauss quadrature of order $n=10,15$ and integrate the system of equations using the RK4 method with adaptive step size \citep{sirca2012} to control the computation precision.

In the limit of high values of the potential $\widetilde\potential$ (i.e~small Roche lobes), we use analytical approximations for the Roche lobe quantities. The analytical expressions are obtained by expressing the radius from the Roche potential in spherical coordinates as a power series of the reciprocal dimensionless potential $\widetilde\potential^{-1}$. The standard formulae for the Roche lobe quantities can then be expressed purely as power series of $\widetilde\potential^{-1}$. \phoebe uses a 9-th degree power series approximation above a threshold value of the potential, which depends on the targeted precision.

\subsection{Rotating stars}

The calculation of volume $V$, derivative of the volume w.r.t.~value of reference potential $V_{,\potential_0}$ and surface area $A$ of rotating stars is simpler than for the Roche lobes because of the rotational symmetry about the $z$ axis. 

We rewrite the isosurface condition $\potential(x,y,z)/\potential_0=1$ using substitutions for Cartesian coordinates $(x,y,z) = \potential_0 (\sqrt{\sigma} \cos\phi, \sqrt{\sigma} \sin\phi, \xi)$ into 
$$
  \frac{1}{\sqrt{\xi^2 + \sigma}}  + \frac{1}{2} b \sigma = 1 \>,
$$
where $b=\omega^2/\potential_0^3$, $\xi\in[-1,1]$ and $\sigma(\xi=\pm 1)=0$. Rotating stellar surfaces are thus given by a single parameter $b$ and analysis shows that closed surfaces only exist for $b \le 8/27$. The volume of the rotating star is given by:
$$
  V = \int_0^1 {\rm d} \xi\,\dot V(\xi)\qquad \mathrm{where} \qquad
  \dot V (\xi) = \frac{2\pi}{\potential_0^3} \sigma(\xi) \>,
$$
while the derivative of the volume w.r.t.~reference value of the potential $V_{,\potential_0} = - \frac{3}{\potential_0} (V + b V_{,b})$ is expressed using:
$$
  V_{,b} = \int_0^1 {\rm d} \xi\, \dot V_{,b}(\xi)\qquad 
  \dot V_{,b}(\xi) = \frac{2\pi}{\potential_0^3} \sigma_{,b}(\xi,\sigma(\xi)) \>,
$$
and the area of the rotating star can be written as:
$$
  A = \int_0^1 {\rm d} t\, \dot A(\xi)\qquad 
  \dot A(\xi) = \frac{4\pi}{\potential_0^2} \sqrt{\sigma(\xi) + {\sigma_{,\xi}(\xi,\sigma(\xi))}^2/4} \>.
$$
The derivatives of $\sigma$ w.r.t.~$t$ and $b$ are:
$$
  \sigma_{,t}(\xi,\sigma) = \frac{2 \xi}{b (\xi^2 + \sigma)^{3/2} - 1}\quad {\rm and} \quad
  \sigma_{,b}(\xi,\sigma) = -\frac{\sigma}{b  - (\xi^2 + \sigma)^{-3/2}} \>,
$$
The above quantities $(V, V_{,b}, A)$ and $\sigma$ form a system of four ordinary differential equations:
$$
 \frac{\rm d}{{\rm d}\xi}(\sigma, V, V_{,b}, A) = (\sigma_{,\xi}, \dot V, \dot V_{,b}, \dot A) \>.
$$
These are integrated over the time interval $\xi\in [0,1]$ where the initial conditions are given at $\xi=1$: $(\sigma, V, V_{,b}, A)(\xi = 1) = 0$ and the resulting surface values are computed at $\xi=0$. The integration in \phoebe is performed with the RK4 method \citep{sirca2012} using adaptive step size for controlled precision. As before, in the limit of small $b$, we use analytic approximations in the form of the 9-th degree power series. For $b<0.1$, the precision of these analytic approximations is better than $10^{-5}$.

\section{The updated reflection model}\label{appendix:reflection_models}

Each component of a binary system has a closed boundary ${\cal M}_i$ and the union of these boundaries is a surface ${\cal M} = \bigcup_i {\cal M}_i$. We define a visibility function for points on this surface as:
$$
  V({\bf r}, {\bf r}') = \left \{
  \begin{array}{lll}
  1&:& \textrm{line of sight } {\bf r}  \leftrightarrow {\bf r}' \textrm{ is unobstructed}\\
  0&:& \textrm{otherwise}
  \end{array}\right. \>,
$$
where ${\bf r}$, ${\bf r}' \in \cal M$. The intensity $I(\hat{\bf e}, {\bf r})$ (i.e.~flux per unit solid angle per projected unit area) of each point on the surface is defined as:
$$
 I(\hat{\bf e}, {\bf r}) = 
 \frac{\partial \Phi}{\partial \potential \partial A \cos \theta}\>,
 \qquad
 \cos\theta =\hat{\bf e} \cdot \hat{\bf n} \>,
$$
where $\hat{\bf e}$ $(\|{\bf e}\|=1)$ is the direction of emission. 

The irradiance $F_{\rm irr}$ (i.e.~radiant flux received by a surface per unit area due to this intensity) at point ${\bf r} \in {\cal M}$ is defined as:
\begin{align}
 F_{\rm irr} ({\bf r}) 
  &= \hat {\cal R} I ({\bf r}) \>, \label{eq:reflection_R} \\
  &= \int_{\cal M} {\rm d} A ({\bf r'}) V({\bf r}, {\bf r}')
  \frac{
  (\widehat{({\bf r}'-{\bf r})}\cdot \hat {\bf n}({\bf r}))
  (\widehat{({\bf r}-{\bf r}')}\cdot \hat {\bf n}({\bf r'}))
  }{|{\bf r} - {\bf r}'|^2} 
  I(\widehat{({\bf r}-{\bf r}')}, {\bf r}') \>,
\end{align}
where the $\hat{}$ designation denotes unit vectors, $\hat {\bf n}$ is the outward pointing normal vector on the surface, and operator $\hat {\cal R}$ maps intensities to irradiances.

If we assume that the intensity $I$ over the surface is described by the Lambert cosine law:
$$
  I(\hat {\bf e};{\bf r}) = I_0({\bf r}) \>,
$$
then the radiant exitance (i.e.~radiant flux emitted by a surface per unit area) at each point is given by:
$$
 F_{\rm ext}({\bf r}) = 
 \int_{\hat {\bf n}\cdot\hat {\bf e}\ge 0} 
 {\rm d} \potential(\hat {\bf e}) I(\hat {\bf e};{\bf r}) 
 = \pi I_0({\bf r})\>.
$$
Using the relation between exitance and intensities, we can express the irradiance with emission described by Lambert's cosine law as:
\begin{equation}
  F_{\rm irr} = \hat {\cal L}_{\rm L} F_{\rm ext}
  \qquad 
  \hat {\cal L}_{\rm L} = \frac{1}{\pi} \hat {\cal R} \>,
  \label{eq:reflection_L_L}
\end{equation}
where we introduce the Lambertian radiosity operator $ \hat {\cal L}_{\rm L}$, which is commonly used in computer graphics (see e.g. \citealt{gershbein1994, cohen1993}). We are now in a position to introduce the limb-darkened intensity as:
$$
  I(\hat {\bf e};{\bf r}) = I_0({\bf r}) 
  D(\hat {\bf e}\cdot \hat {\bf n}; {\bf r} ) \>,
$$
where $I_0({\bf r})$ is the normal emergent intensity and $D$ is the limb-darkening factor: $D(1; {\bf r}) =1$. The corresponding radiant exitance then becomes:
$$
  F_{\rm ext}({\bf r}) = 
  \int_{\hat {\bf n}\cdot\hat {\bf e}\ge 0} {\rm d} \potential(\hat {\bf e})  I({\bf e}; {\bf r}) \hat{\bf e}\cdot \hat {\bf n}
  =  I_0({\bf r}) D_0({\bf r}),\qquad
  D_0({\bf r}) = \int_{\hat {\bf n}\cdot\hat {\bf e}\ge 0} 
  {\rm d} \potential(\hat {\bf e}) 
  D(\hat{\bf e}\cdot \hat {\bf n}; {\bf r} ) \hat{\bf e}\cdot \hat {\bf n}.
$$
We can again express the irradiation as a function of the radiant exitance by introducing the limb-darkened radiosity operator $\hat {\cal L}_{\rm LD}$:
\begin{equation}
  F_{\rm irr} = \hat {\cal L}_{\rm LD} F_{\rm ext}
  \qquad 
  \hat {\cal L}_{\rm LD} = \hat {\cal R} \circ \frac{D}{D_0} \>.
  \label{eq:reflection_L_LD}
\end{equation}
Both the irradiation and exitance are non-directional fluxes per unit area and with the two radiosity operators, ${\cal L}_{\rm L}$ and ${\cal L}_{\rm LD}$, we are able to elegantly express the irradiation as a function of exitance. 

Using the notation of radiosity operators we can introduce a simple implementation of reflection models in \phoebents: the model of \citet{wilson1990} and the approach presented here. The goal of all reflection models is to calculate the total radiosity $F_{\rm out}$ of a surface, given the intrinsic exitance $F_0$ or intensity $I_0$ of each surface point.

Wilson's model assumes that the total radiosity $F_{\rm out}$ is composed of the intrinsic emittance and reflected irradiance given as radiosity diffusely radiated from the surface using the limb-darkening law:
\begin{equation}
  F_{\rm out}({\bf r}) = 
  F_{\rm 0}({\bf r}) + \rho({\bf r}) \hat{\cal L}_{\rm LD} F_{\rm out}({\bf r}) \>.
  \label{eq:reflection_wilson}
\end{equation}
%

Strictly speaking, the justification for this treatment of reflected light is not warranted. Because heating by the absorbed light is neglected, the reflected and the intrinsic light are thus not constrained and have to be accounted for separately. As in Wilson's model, the radiosity $F_{\rm out}$ is composed of intrinsic emittance $F_0$ and irradiance $F_{\rm in}$ reflected from the surface. In our model, the irradiance is given as intrinsic emittance following the limb-darkening law and irradiance diffussed with the Lambertian cosine law. The equations determining the model are:
\begin{align}
  &F_{\rm in}({\bf r}) = 
  \hat{\cal L}_{\rm LD} F_0({\bf r}) + \hat{\cal L}_{\rm L} (\rho F_{\rm in})({\bf r}) \>,
   \label{eq:reflection_horvat}\\
  &F_{\rm out}({\bf r}) = F_0({\bf r}) + \rho({\bf r}) F_{\rm in}({\bf r}) \>. \label{eq:radiosity_ours}
\end{align}
In this approach, the bodies act as limb-darkened radiators of intrinsic light and ideal Lambertian diffusive radiators of incoming light. In the case where both radiosity operators are the same, ${\cal L}_{\rm LD} = {\cal L}_{\rm L} := {\cal L}$, expression (\ref{eq:radiosity_ours}) reduces to Wilson's Eq.~(\ref{eq:reflection_wilson}).

In order to use these reflection models in practice, we need to discretize the operators and solve the corresponding equations. We start by partitioning $\cal M$ into disjoint parts:
$$
  {\cal M} = \bigcup_i {\cal S}_i\qquad {\cal S}_i \cap {\cal S}_j = 0\qquad i\neq j \>.
$$
The introduced radiosity operators are linear and have the following functional form:
$$
  \hat {\cal L}_* f({\bf r}) = 
  \int_{\cal M} {\rm d} A({\bf r'}) K_*({\bf r}, {\bf r}') f({\bf r}')\qquad
  *  = {\rm L}, {\rm LD} \>.
$$
where the kernel $K_*$ can refer to the Lambertian or limb-darkening model. Assuming that all functions are constant over each element ${\cal S}_i$ of the partition:
$$
  f({\bf r}) = \sum_i f_i \chi_i ({\bf r})  \qquad
  \chi_i ({\bf r}) = \left \{
  \begin{array}{lll}
  1 &:& {\bf r} \in {\cal S}_i \\
  0 &:& \textrm{otherwise}
  \end{array}
  \right. \>.
$$ 
The $\chi_i$ is a characteristic function over ${\cal S}_i$ and the set of all $\{\chi_i \}_i$ functions is our functional basis. The relation $f' = \hat {\cal L}_* f$ where the functions $f$ and $f'$ are expanded in the new basis with coefficients $f_i$ and $f_j$, respectively, can be written as:  
$$
  f_i' = \sum_j L^*_{i,j} f_j 
$$
with
$$
  A_i = \int_{{\cal S}_i} {\rm d} A({\bf r})
  \qquad
  L^*_{i,j}= \frac{1}{A_i}
  \int_{{\cal S}_i} {\rm d} A({\bf r}) 
  \int_{{\cal S}_j} {\rm d} A({\bf r'}) K_*({\bf r}, {\bf r}') \>.
$$
The coefficients $L_{i,j}^*$ are generalizations of the view-factors \citep{modest2003} and by collecting them in a matrix ${\bf L^*} = [L^*_{i,j} ]_{i,j}$ we get a discretized version of the operator $\hat {\cal L}_*$. 

The working surface $\cal M$ in \phoebe is the mesh of triangles built over the smooth isosurface of an astrophysical object. We consider two approaches to discretizing operators: per-triangle discretization and per-vertex discretization. In both cases we simplify the expressions for generalized view-factors:
$$
 L_{i,j}^* \approx 
 A_j K_*({\bf r}_i, {\bf r}_j) 
$$
where ${\bf r}_i$ and ${\bf r}_j$ are characteristic points chosen depending on the discretization scheme and $A_i$ are their attached areas. In per-triangle discretization, ${\bf r}_i$ are the triangle barycenters, ${\bf n}_i$ are the triangle normals and $A_i$ are the triangle areas, whereas in per-vertex discretization, ${\bf r}_j$ are the vertices, ${\bf n}_i$ are the vertex normals and $A_i$ is $1/3$ of the sum of triangle areas sharing each vertex. 

\section{Horizons on isosurfaces}\label{appendix:horizon}

The boundaries of our astrophysical bodies (binary stars, rotating stars, etc.) are defined as isosurfaces of the corresponding scalar potential function $\Phi$. If we observe a smooth isosurface at a reference value $\Phi_0$:
\begin{equation}
  \Phi({\bf r}) = \Phi_0,\qquad {\bf r} \in \mathbb{R}^3 \>
  \label{eq:horizon_pot}
\end{equation}
 from direction $\hat {\bf v}$ ($\|{\hat{\bf v}}\|$=1), the horizon is a 1-dimensional manifold that satisfies the equation:
\begin{equation}
  {\bf g}({\bf r})\cdot {\hat{\bf v}} = 0 \>, 
  \label{eq:horizon_orto}
\end{equation}
where ${\bf g} = \nabla \Phi$ is the gradient of the potential. The horizon of a smooth surface is a finite set of closed curves, which divides the isosurface into visible and invisible parts with respect to the external observer.

Let ${\bf r}(s)$ represent a naturally parametrized curve of the horizon. Its variation over the surface is given by the following differential equation:
\begin{equation}
  \frac{{\rm d}{\bf r}}{{\rm d} s} 
  = \frac{({\bf H}({\bf r})\cdot {\hat{\bf v}}) \times {\bf g}({\bf r})}               
         {\|({\bf H}({\bf r})\cdot {\hat{\bf v}}) \times {\bf g}({\bf r}) \|} \>,
  \label{eq:horizon_ODE}
\end{equation}
where ${\bf H}({\bf r})= (\nabla\otimes\nabla) \Phi ({\bf r})$ is the Hessian matrix of the potential $\Phi$. 

The curve of the horizon is obtained by integration of the differential equations (\ref{eq:horizon_ODE}) from a starting point ${\bf r}^*$ using the RK4 method. The starting point for the integration is obtained by solving the system of equations (\ref{eq:horizon_pot}) and (\ref{eq:horizon_orto}) via a Newton-Raphson iteration:
$$
  {\bf r}_{k+1} = {\bf r}_k + \delta {\bf r}({\bf r}_k) \>.
$$
The linear expansion of the equations, as requried by the Newton-Raphson scheme, reveals that the iteration step $\delta {\bf r}$ is uniquely determined. Its minimal length variant is given by:
$$
 \delta {\bf r}({\bf r})  = 
 \alpha\, {\bf g}({\bf r}) + \beta\,{\bf H}({\bf r})\cdot\hat{\bf v} \>,
$$
where the weights $\alpha$ and $\beta$ are calculated via:
$$
 \left[
  \begin{array}{c}
  \alpha \\
  \beta 
  \end{array}
\right] = 
\left[
  \begin{array}{cc}
  \|{\bf g}({\bf r})\|^2 & {\bf g}({\bf r})\cdot{\bf H} ({\bf r})\cdot\hat{\bf v} \\
  {\bf g}({\bf r})\cdot{\bf H}({\bf r})\cdot\hat{\bf v} & \|{\bf H}({\bf r})\cdot\hat{\bf v}\|^2
  \end{array}
\right]^{-1}
\left[
  \begin{array}{c}
  \Phi_0 - \Phi({\bf r}) \\
  - {\bf g}({\bf r}) \cdot \hat{\bf v}
  \end{array}
\right] \>.
$$
The iteration is initiated from a point ${\bf r}_0$ that lies sufficiently inside the isosurface, but sufficiently far so that the constraining  of the inverse does not have a profound effect.

\end{document}